\documentclass[fleqn,usenatbib]{mnras}

\usepackage{newtxtext,newtxmath}

\usepackage[T1]{fontenc}

\DeclareRobustCommand{\VAN}[3]{#2}
\let\VANthebibliography\thebibliography
\def\thebibliography{\DeclareRobustCommand{\VAN}[3]{##3}\VANthebibliography}

\usepackage{graphicx}	
\usepackage{amsmath}	

\title[PLATOSpec instrument for the community]{PLATOSpec: a precise spectrograph in support of space missions}

\author[P. Kabáth et al.]{
P. Kabáth,$^{1}$\thanks{E-mail: kabath@asu.cas.cz}
M. Skarka,$^{1}$
A. Hatzes,$^{2}$
E. Guenther,$^{2}$
L. Vanzi,$^{3}$
R. Brahm,$^{4}$
J. Janík,$^{5}$
P. Pintr,$^{6}$
P. Gajdoš,$^{1,7}$\newauthor
J. Lipták,$^{1,8}$
J. Žák,$^{1}$
H. M. J. Boffin,$^{9}$
L. Antonucci,$^{3}$
G. Avila,$^{13}$
Z. Balkóová,$^{1,8}$
M. E. Ball,$^{2}$
M. Flores,$^{3}$\newauthor
A. Fuentes,$^{3}$
J. Fuchs,$^{1}$
R. Greimel,$^{11}$
A. Gajardo,$^{4}$
V. D. Ivanov,$^{9}$
J. K\"ohler,$^{2}$
M. Leitzinger,$^{10}$\newauthor
T. Moravčík,$^{1,8}$
J. Nečásek,$^{6}$
R. J. U. Neubert,$^{2}$
P. Odert,$^{10}$
G. Olguin,$^{3}$
M. Tala Pinto,$^{4,12}$
M. Roth,$^{2}$\newauthor
L. Řezba,$^{1}$
V. Schaffenroth,$^{1}$
M. Sigwarth,$^{2}$
J. Srba,$^{1}$
A. Suárez,$^{3}$
P. Škoda,$^{1}$
J. Šubjak,$^{1}$
J. Václavík,$^{6}$
M. Veselý,$^{6}$\newauthor
R. Veselý,$^{1}$
M. Vítková,$^{1,5}$
J. U. Winkler,$^{2}$
M. Zummer,$^{1,8}$
E. Ždárská,$^{1}$
\\
$^{1}$ Astronomical Institute of the Czech Academy of Science, Fričova 298, 251 65 Ondřejov, Czech Republic\\
$^{2}$ Thüringer Landessternwarte, D-07778 Tautenburg, Germany\\
$^{3}$ Center of Astro Engineering, Pontificia Universidad Católica de Chile, Av. Vicuña Mackenna 4860, 782-043 Santiago, Chile\\
$^{4}$ Faculty of engineering and sciences, Universidad Adolfo Ibañez, Av. Diag. Las Torres 2640, 7941169 Santiago, Chile\\
$^{5}$ Department of Theoretical Physics and Astrophysics, Masaryk University, Kotl\'a\v{r}sk\'a 2, CZ-611 37 Brno, Czech Republic\\  
$^{6}$ Institute of Plasma Physics of the Czech Academy of Sciences, Research Centre for Special Optics and Optoelectronic Systems TOPTEC,\\ 
 U Slovanky 2525/1a 182 00, Praha 8 \\  
$^{7}$ Institute of Physics, Faculty of Science, Pavol Jozef \v{S}af\'arik University, Park Angelinum 9, 040 01 Ko\v{s}ice, Slovakia\\
$^{8}$ Astronomical Institute of Charles University, V Hole\v{s}ovi\v{c}k\'{a}ch 2, CZ-180 00 Prague, Czech Republic\\
$^{9}$ European Southern Observatory, Karl-Schwarzschild-str. 2, 85748 Garching, Germany\\ 
$^{10}$ Institute of Physics, Department for Astrophysics and Geophysics, University of Graz, Universit\"{a}tsplatz 5, A-8010 Graz, Austria\\
$^{11}$ RG Science, Schanzelgasse 17, A-8010 Graz, Austria\\
$^{12}$ Department of Astronomy, The Ohio State University, 140 W. 18th Ave., Columbus, OH, 43210 \\  
$^{13}$ Baader Planetarium GmbH, Zur Sternwarte 4, D-82291, Mammendorf, Germany \\
 } 

\date{Accepted XXX. Received YYY; in original form ZZZ}

\pubyear{\the\year{}}

\begin{document}
\label{firstpage}
\pagerange{\pageref{firstpage}--\pageref{lastpage}}
\maketitle
\begin{abstract}
The upcoming space missions that will characterize exoplanets, such as PLATO and Ariel, will collect huge amounts of data that will need to be complemented with ground-based observations. The aim of the PLATOSpec project is to perform science with an echelle spectrograph capable of measuring precise radial velocities. The main focus of the spectrograph will be to perform the initial screening and validation of exoplanetary candidates, in addition to study stellar variability. It will be possible to determine the physical properties of large exoplanets. The PLATOSpec blue-sensitive spectrograph, with a spectral range of 380 to 700\,nm and a resolving power of $R=70,000$, is installed on the 1.5-m  telescope at the ESO La Silla Observatory in Chile. Initial results show that the radial-velocity limit given by the wavelength calibration is about 2-3\,m/s. Tests on bright F-K main-sequence standard stars reveal a scatter of about 5\,m/s over a few hours. The scatter over a few months is  slightly higher. We demonstrate the capabilities of PLATOSpec on the mass determination of WASP-79\,b and the spin-orbit alignment of WASP-62\,b via the Rossiter-McLaughlin effect. We show its possible usage on variable star research as demonstrated on the false-positive exoplanetary candidate TIC\,238060327, which is proven a binary star. Investigation of line-profile variations of the roAp star $\alpha$\,Cir shows that PLATOSpec can also be used  for the surface mapping. Finally, we present new results on the active star UY Pic in the PLATO southern field. Our results show that PLATOSpec is a versatile spectrograph with great precision.
\end{abstract}




\section{Introduction}

The first space mission dedicated to the exploration of exoplanetary systems, \textit{CoRoT} \citep{2009A&A...506..411A}, demonstrated that an extremely thorough candidate validation process from the ground needs to be established. 
Later, \textit{Kepler} provided improved statistics, reporting a non-negligible false positives rate 
\citep{2013ApJ...766...81F}, 
with different types of planets having different rates, depending on their orbital periods and masses \citep{2016A&A...587A..64S, 2011ApJ...738..170M,2012MNRAS.426..342C}. Therefore, for the TESS mission \citep{2015JATIS...1a4003R}, a new concept of dedicated ground-based follow-up -- TFOP\footnote{\url{https://tess.mit.edu/followup/}} \citep[e.g.,][]{Collins2019,Schroll2025} -- was introduced. Various observatories joined the TFOP group using their facilities for a specific part of the follow-up. Work distribution between smaller observatories with mid-aperture telescopes and large facilities made the process indeed much more efficient. The candidate vetting process was streamlined and only excellent candidates were passed on for precise mass determination. 
Furthermore, several consortia dedicated to the validation and characterization of planetary candidates, such as KESPRINT\footnote{\url{https://kesprint.science}}\citep[e.g.,][]{Barragan2016,Kabath2022,Cabrera2023} or WINE\footnote{\url{https://sites.google.com/view/wine-exoplanets/home}}\citep[e.g.,][]{Brahm2019,Brahm2020,Pinto2024} were founded. 
Such consortia that own telescope time can moreover play an increasing role in long-term monitoring projects, such as the detection of planets on long orbits \citep[e.g.,][]{Trifonov2023,Vitkova2025}, stellar variability monitoring for space missions, and stellar activity observations \citep[e.g.,][]{Wollmann2023,2025MNRAS.537..537O}. 
For the upcoming PLATO space mission, the false positive rate predictions vary from a few percent up to 60\%, based on the stellar and planetary radii \citep{2023MNRAS.518.3637B,2025MNRAS.540.2578B}, highlighting again the importance of a thorough ground-based vetting system.

The PLATO space mission -- dedicated to the hunting of earth-sized planets on long orbits \citep{Rauer2025}  -- will be launched at the end of 2026. The main PLATO target sample is selected so that it can be followed-up from the ground \citep{Montalto2021} and a dedicated team -- the GOP team -- is now developing the strategies to efficiently validate PLATO candidates. The follow-up process is split into various parts, such as the preliminary screening of stellar parameters, the determination of false positives, the monitoring of stellar activity, and, finally, the determination of planetary masses via precise radial velocity measurements.
For each of these parts, different instruments and techniques are required. The most versatile instrumentation which can cover many parts is a precise spectrograph with a high resolving power and installed on a 2--4-m class telescopes. Successful examples of such instruments are ELODIE \citep{1996A&AS..119..373B}, CORALIE \citep{2010A&A...511A..45S,2013A&A...551A..90M}, TRES \citep{Szentgyorgyi2007}, OES \citep{2020PASP..132c5002K}, TCES \citep{2003ESASP.539..441H}, FEROS \citep{1999ASPC..188..331S}, as well as HARPS \citep{2003Msngr.114...20M,2000SPIE.4008..582P}, FOCES \citep{refId0}, MaHPS and SOPHIE \citep{2008SPIE.7014E..0JP}, AAT with UCLES \citep{fd} and VELOCE \citep{2024SPIE13096E..45T}, HARPS-N \citep{2012SPIE.8446E..1VC} and recently NIRPS \citep{2024SPIE13096E..0CA}. The latter spectrographs are placed in vacuum vessels and therefore capable of ultra-precise measurements of radial velocities down to the cm/s regime. 


Despite this impressive armada, with the data avalanche foreseen by PLATO, more spectrographs are needed for planetary candidate vetting processes, and, if possible, for monitoring stellar variability -- thus requiring enhanced sensitivity in the Ca~\textsc{ii} H\&K wavelength range. Furthermore, deriving the stellar parameters, temperature ($T_{\rm eff}$), gravity ($\log~g$), and metallicity ([Fe/H]), is adequately done with mid-aperture-sized instrumentation. When combined with the asteroseismological data from PLATO and Gaia data, these parameters allow the precise determination of the masses and radii of the host stars. 

In this paper, we present the early science done with a new spectrograph, PLATOSpec, which was installed in November 2024 at the ESO 1.52-m telescope in La Silla,  Chile. PLATOSpec is a precise thermally stable echelle spectrograph capable of measuring radial velocities down to about 3\,m/s. PLATOSpec has enhanced sensitivity in the blue, so that it can observe lines important for monitoring of the stellar variability. PLATOSpec will contribute to the follow-up efforts of the space missions PLATO and Ariel by providing stellar characterization, planetary mass, and eccentricity, as well as the projected spin-orbit angles of the planetary orbits \citep{zak25ps}. It is already involved in the TESS mission follow-up, with first results presented here. 

The PLATOSpec consortium consists of three main partners: the Astronomical Institute of the Czech Academy of Sciences, Czech Republic; Thueringer Landessternwarte Tautenburg, Germany; and Pontifica Universidad Católica de Chile, Chile; and three minor partners: Masaryk University, Czech Republic; Universidad Adolfo Ibanez, Chile; and the Institute for Plasma Physics, Czech Republic. 


\section{The PLATOSpec project}\label{Sect:Project}


The PLATOSpec project \citep[][]{Kabath2022a,Kabath2024} includes the refurbishment of the ESO 1.52-m telescope\footnote{https://www.eso.org/public/teles-instr/lasilla/152metre/} (E152), followed by the construction and installation of the PLATOSpec instrument itself. 

\subsection{The E152 telescope}\label{Sect:Telescope}

Since its inauguration in 1967, the E152 telescope was hosting various instruments such as the Coud\'e spectrograph, the Boller \& Chivens, and FEROS. The telescope was decommissioned in 2003. In April 2022, after a successful upgrade of the control system enabling remote and weatherproof operations, the telescope was re-inaugurated. In the same year, a test instrument, the echelle spectrograph PUCHEROS+ \citep{2012MNRAS.424.2770V,Kabath2024,10.1093/mnras/staf1290}, was installed at the E152, collecting useful scientific data over 2023 and 2024 \citep{2025MNRAS.537..537O,2024A&A...688A.120S,2025AJ....169...47K}. The PLATOSpec spectrograph was finally installed and commissioned for science on 26 November 2024.

The operations of the telescope are controlled from remote locations via custom-made GUIs and daemon services.  Routine observing can be done semi-automatically via the scripting of chains of targets. However, the observer still needs to confirm the telescope pointing during the acquisition. The automatic observing targets database, scheduler, and observing night-log system were all developed at the Astronomical Institute of the Czech Academy of Sciences in Ondřejov (AIASCR) and are based on {\tt astropy} \citep{astropy:2013,astropy:2018,astropy:2022} and {\tt astroplan} routines \citep{Morris2018}. 

The telescope has a comprehensive weather and safety monitoring system. In case of a 
network connection failure, it can operate autonomously until it is closed either due to the sunrise or due to any technical problem. Currently, human guidance is still needed for the pointing procedure; however, an upgrade towards a fully automated process is expected soon. 

\begin{figure}
\includegraphics[width=0.49\textwidth]{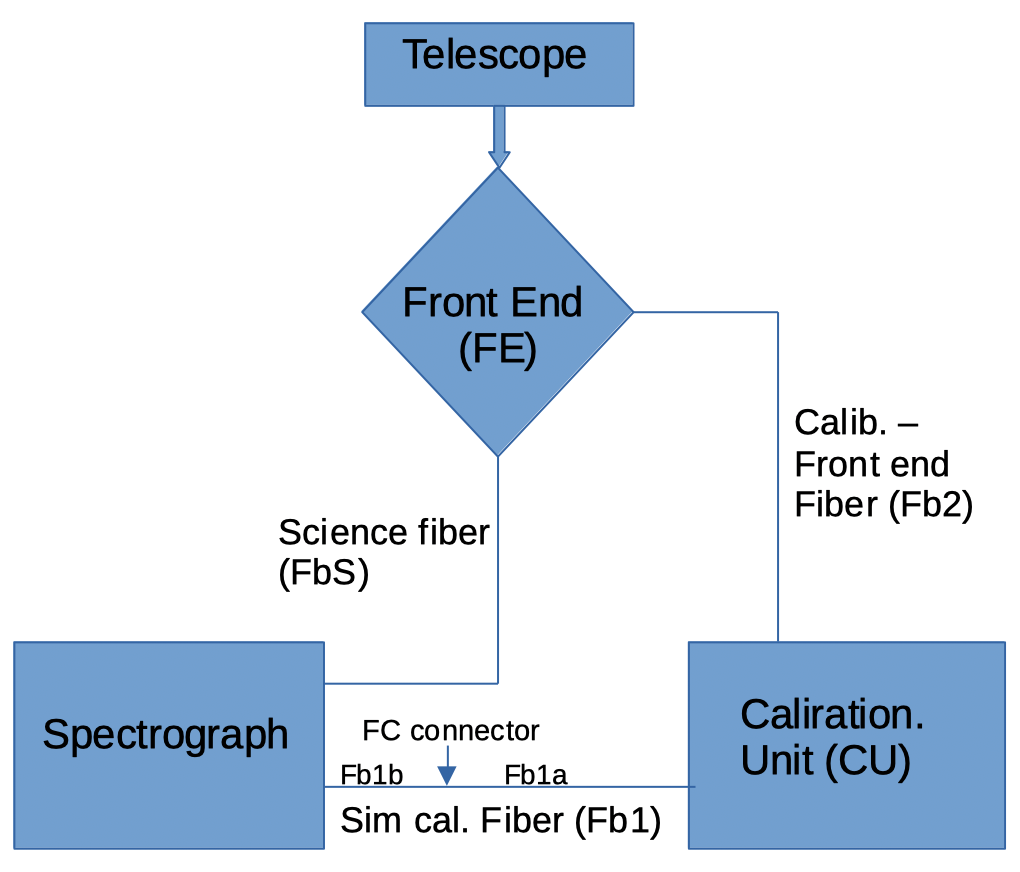}
\caption{Light path from the telescope to the spectrograph (FbS) via the front end and the fiber linking the calibration unit to the spectrograph (Fb1) and the front end (Fb2).} 
\label{block}
\end{figure}

A block diagram illustrating the light path between each component can be found in Fig.\,\ref{block}.

\subsection{The front end}\label{Sect:FE}

\begin{figure}
\includegraphics[width=0.49\textwidth]{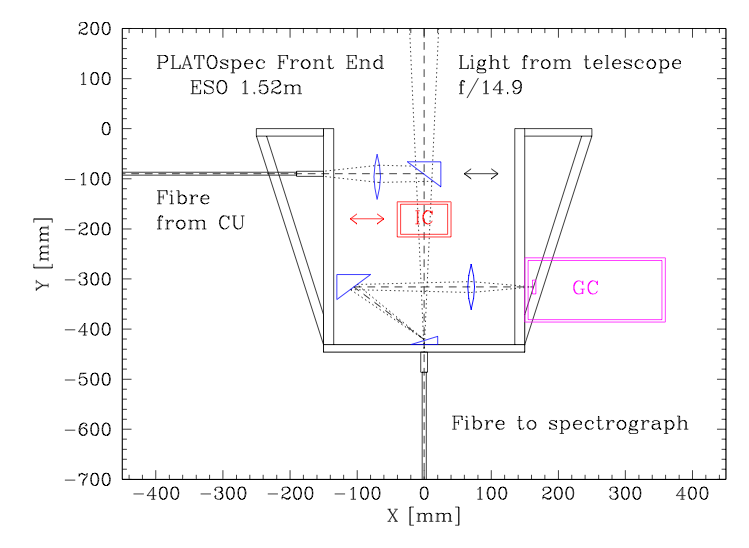}
\caption{Light path in the front end. At the bottom, a set of mirrors reflects light to the guiding camera, while a movable mirror (blue triangle in the beam from the telescope) can reflect light from the calibration unit to the spectrograph through the single scientific fiber. The front end includes also an iodine cell (IC) that can be moved in the beam if requested. 
} 
\label{fig:fe}
\end{figure}

The telescope's front end (FE; Fig.~\ref{fig:fe}) is installed in the Cassegrain focus and includes the guiding system and a fiber injection system. Since PLATOSpec is not a vacuum stabilized spectrograph for precise radial velocities measurements, we chose to use an iodine absorption cell (located in the front end and in the calibration unit) to provide the wavelength calibration for Doppler measurements. The iodine cell is movable and can be used when observing with the science fiber. 

The guiding is performed with a CMOS detector-equipped camera (C1+12000), delivered by Moravian Instruments\footnote{https://www.gxccd.com}, and with a field of view of about 4.6$\times$3.4 arcminutes. The guiding is performed on the science fiber by a custom written automatic routine. Typically, the corrections are performed every 5-10 seconds, depending on the brightness of the guide star. 
We further plan to install a tip-tilt glass plate for fine guiding at the end of 2025.

The fiber injection system consists of a pinhole in front of a triplet lens glued on the fiber. The pinhole has a diameter of 0.25 mm, which corresponds to 2\arcsec on the sky. The original light beam from the telescope with foal ratio of F$/14.9$ is converted to F$/4.5$ and injected to the octagonal fiber with a diameter of 50 microns.  

\subsection{The calibration unit}\label{Sect:CU}

\begin{figure}
\includegraphics[width=0.49\textwidth]{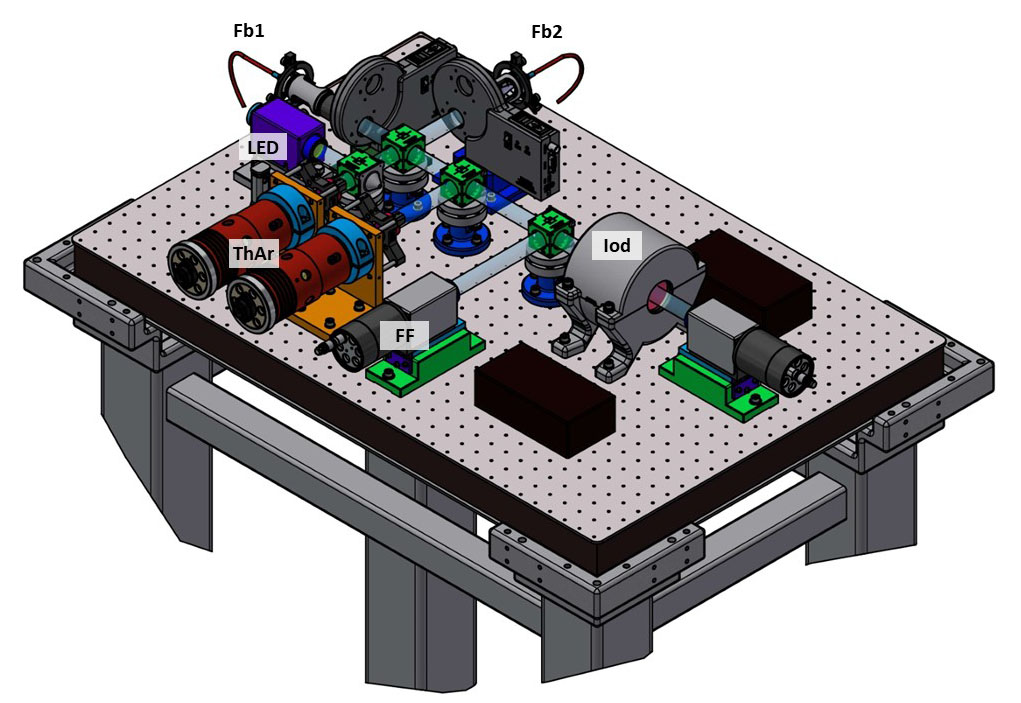}
\caption{The calibration unit consists of two ThAr lamps and, for simultaneous calibration, a flat lamp (FF) with an additional UV-LED (LED) source and an iodine cell (Iod). It is connected by another optical fiber to the front end (Fb2) and by other fiber (Fb1) to the spectrograph.}
\label{fig:calu}
\end{figure}

The calibration unit (Fig.~\ref{fig:calu}) is located in a separate room. It is connected to spectrograph with an optical fiber Fb1 in Fig.~\ref{fig:calu} and with another fiber, Fb2 in Fig.\,\ref{fig:calu} connecting it to the Front end. The diameter of the Fb2 fiber is 400 $\mu$m. The calibration unit to spectrograph fiber Fb1 has a total length of 8.5 meters and it consists of two parts Fb1a and Fb1b. The ongoing 8 meter long Fb1a part from the calibration unit is a 100\,$\mu$m circular fiber that connects through an FC connector into a 0.5 meter long Fb1b octagonal 50\,$\mu$m calibration fiber. The octagonal calibration fiber Fb1b joins the scientific octagonal fiber FbS incoming to the spectrograph directly from the front end. The fiber injection into the spectrograph is performed with a fiber holder fixing the scientific and calibration fibers separated by 54 $\mu$m. 

The calibration unit is equipped with two lamps used for flat fielding -- an UV-LED and a continuum lamp -- that can be used simultaneously. There are also two Thorium-Argon (ThAr) lamps, where one is used as a backup. Furthermore, another iodine cell can be used for simultaneous (with science exposure) calibration instead of the ThAr lamp, so that the signal from the cell is injected into the calibration to spectrograph fiber. The IC is illuminated with a white lamp operating in the range of 520-620 nm. This enables us to record the iodine spectrum adjacent to the stellar spectrum and not on the target' spectrum, which is the normal procedure when using an iodine cell.  
To our knowledge this is the only instrument using simultaneous iodine cell calibration and we discuss our experience with this mode below.

Thus, the calibration unit provides the following observing modes:
\begin{itemize}
\item Simultaneous ThAr calibration through the calibration fiber (through Fb1)
\item ThAr calibration through the front end and science fiber (through Fb2)
\item Simultaneous iodine cell calibration through the calibration fiber (through Fb1)
\item Front end iodine cell calibration (in-situ) through the science fiber 
\item Flat field calibrations through the calibration fiber (through Fb1)
\item Flat field calibration through the science fiber through the front end (through Fb2)
\end{itemize}

The calibration unit is equipped with a separate control and supply unit and can be operated from the local terminal or remotely.

\begin{figure}
\includegraphics[width=0.49\textwidth]{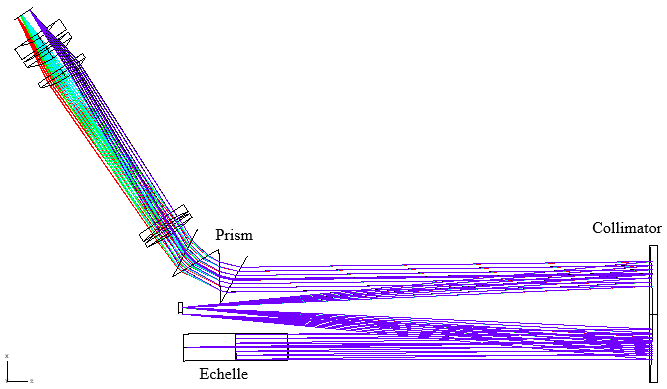}
\caption{Optical design of the elements of the spectrograph.}
\label{fig:spec}
\end{figure}

\subsection{The spectrograph}\label{Sect:SpectDescription}

PLATOSpec is a white pupil echelle spectrograph with spectral resolving power of $R=70,000$, covering the range of wavelengths from 380 to 700\,nm over 55 spectral orders. The spectrograph layout is depicted in Fig.~\ref{fig:spec}. 

The light is injected into the spectrograph from a 50 $\mu$m core octagonal fiber from the front end (scientific fiber)  and from a similarly sized calibration fiber from the calibration unit (calibration fiber). The focal ratio of the incoming beam from the scientific fiber is converted from F/4.5 to F/18 on the entrance to the spectrograph.
The light beam first hits the image slicer slicing the beam into 2 slices, and is then collimated by the parabolic mirror, which reflects the light to an echelle grating. From the grating, the light beam travels back to the collimator, which sends it to the cross-disperser prism and subsequently to the objective. The detector is an ANDOR Ikon GLx with 2048x2048 pixels and optimized for the blue wavelength ranges thanks to a special coating. The detailed optical design description will be provided in technical papers which are currently in prep. All important parameters of the spectrograph are summarised in Tab.\,\ref{tsp}


\begin{table}
	\centering
	\caption{PLATOSpec essential parameters summary.}
	\label{tsp}
	\begin{tabular}{lc} 
		\hline
		Parameter & Value  \\
		\hline
        Spectrograph feeding & fiber, image slicer (2 slices) \\
        Spectrograph type & echelle, white pupil \\
        Detector & Andor Ikon GLx (2k$\times$2k) \\
		Wavelength range & 380-700 (nm) \\
        Resolving power & 70,000 \\
        Number of orders & 55 \\
        Sampling (pix) & 2 \\
		Calibration & sim ThAr, sim Iodine\\
        Precision in radial velocities & 3-5 m/s\\
		\hline
	\end{tabular}
\end{table}

\subsection{On-sky performance}\label{Sect:Performance}

\begin{figure}
\includegraphics[width=0.495\textwidth]{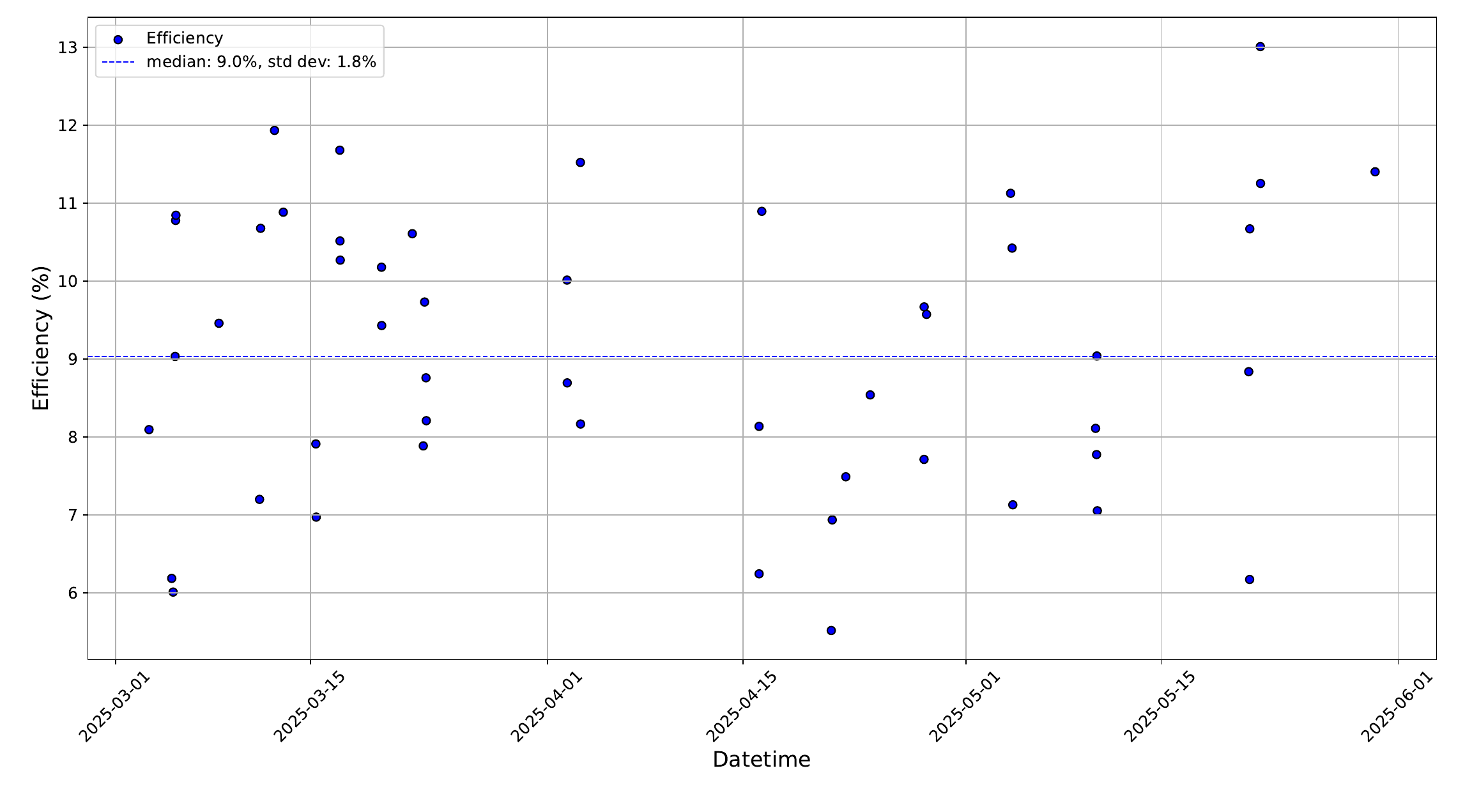}
\caption{Sample of optimal measurements used for the efficiency determination. The median value of 9\% is represented by the blue dashed line. }
\label{fig:eff}
\end{figure}

We performed a few tests to determine the overall efficiency of the spectrograph and to derive the signal-to-noise ratio (SNR) versus exposure time performance. For the tests, a sample of 50 observations obtained between March, 2025 and June 10, 2025 was used. The airmass of the stars was always better than 1.1 and the seeing proxy determined on the guider CCD was always better than 4\arcsec. The seeing on the guider is about 1-2\arcsec systematically higher than the new RINGSS seeing value reported by La Silla observatory since January 2025 \citep{10.1093/mnras/staa4049}. However, we checked and there is no apparent dependence of efficiency on the seeing. From our measurements (Fig.~\ref{fig:eff}), we derive a median efficiency of the whole system (telescope plus spectrograph) of 9\%.  

\begin{figure}
\includegraphics[width=0.495\textwidth]{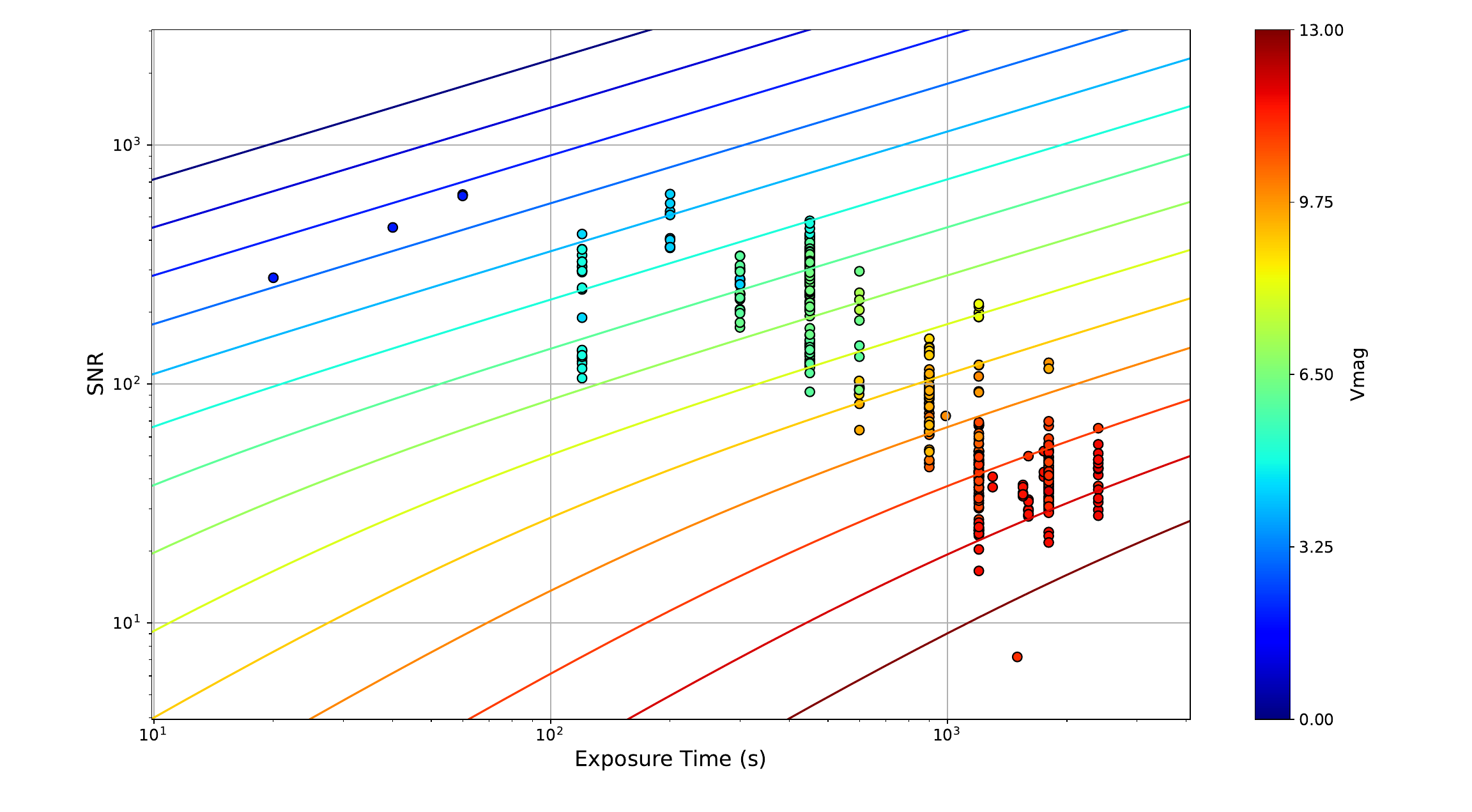}
\caption{SNR vs. exposure time performance for a sample of stars. The overplotted lines roughly correspond to the expected SNR for a given exposure time and a given magnitude of the star, the latter being represented by the colour bar. }
\label{fig:snrv}
\end{figure}

For the exposure time calculator, we used objects of various  brightnesses and determined the SNR for each exposure. A graph of SNR vs. exposure time for different stars is shown in Fig.~\ref{fig:snrv}. As a further test, we also looked at the faintest star in our sample, which had  $V=13.7$\,mag. In an 1800s exposure, an SNR of six was obtained. 

We provide also some extensive discussion of the effect of atmospheric dispersion on the performance of the spectrograph in  Appendix~\ref{diffdis}.

\subsection{Simultaneous photometry}\label{Sect:SimPhot}

In addition to the main telescope with the spectrograph, there are two attached finder scopes, both equipped with 15-cm lenses (f/10, focal length 1530 mm). One of the finder scopes is equipped with a Moravian Instruments C4-16000 CMOS camera, allowing us to do fast photometry simultaneously with the spectroscopic observations. The camera has also a filter wheel with Sloan \textit{ugriz}, narrow-band H$\alpha$, and wide blue-band filters. The field of view of this photometric setup is about  1.4$\times$1.4 degrees.

The other finder telescope is fitted with a ZWO ASI 2600MM-PRO camera utilizing a Sony IMX571 CMOS detector. The eight position filter wheel contains Sloan \textit{ugri} and Johnson $B,V$ filters, all from Baader, a Paton Hawksley SA 200 grating and an empty slot. It has a plate scale of 0.5 arcsec/pixel and a field of view of 52x35 arcmin. The spectral resolution for slitless spectroscopy is strongly affected by chromatic aberration from the achromatic refractive finder telescope, varying between $R\approx24$ to $R\approx200$. To focus the images and to be able to rotate the field of view for slitless spectroscopy, a Primalucelab ESATTO 2\arcsec focuser and an ARCO 2\arcsec rotator are used.

We do not present any scientific case in the Section\ref{Sect:FirstResults} but it was proven elsewhere\citep{2025MNRAS.537..537O} that especially for the monitoring of stellar variability for brighter sources, the photometry is useful and reaching to about 0.025 mag for roughly 9 magnitude star.

\section{The data flow}\label{Sect:DataFlow}

The raw data are initially stored in the computer of the control room at La Silla. These data are also directly copied to our archive at the Astronomical Institute in Ond\v{r}ejov from where they are available for the community after a proprietary period which is typically one year for most of the data. After the data processing performed on a separate computer (see below), the reduced spectra and other data products are available from the open archive in Ond\v{r}ejov. Starting in 2026, the data will be offered also from the ESO archive, as is usual for ESO instrumentation, making the impact for the community even broader and the data more accessible.

\subsection{Simultaneous ThAr calibration}

PLATOSpec data obtained with a ThAr lamp as a simultaneous calibration source are processed with a dedicated and automated pipeline based on a new version of the \texttt{ceres} package \citep{ceres}, named \texttt{ceres+}. This is an object-oriented \texttt{Python} module with a wide variety of algorithms to process, reduce, calibrate, and analyse data obtained from echelle spectrographs with different specifications.

Briefly, the pipeline for PLATOSpec works over all the images in a directory corresponding to a given night. The first step is to classify the images according to the information available in the header. The types of images used in the processing are:
\begin{itemize}
    \item \textit{bias}: CCD bias frame.
    \item \textit{flat-sci}: Image where the science fiber is illuminated by a continuum lamp.
    \item \textit{flat-cal}: Image where the calibration fiber is illuminated by a continuum lamp. If needed, flat fields in science and calibration fibers can be illuminated simultaneously.
    \item \textit{ThAr}: Image where both fibers contain light of the ThAr lamp.
    \item \textit{science-sim}: Image where the science fiber is illuminated by a star and the calibration fiber is illuminated by the ThAr lamp.
    \item \textit{science-nosim}: Image where the science fiber is illuminated by a star and there is no light in the calibration fiber.
\end{itemize}
All \textit{bias}, \textit{flat-sci}, \textit{flat-cal} are combined by type to generate a master calibration frames: \textit{masterbias}, \textit{masterflat-sci}, \textit{masterflat-cal}. The masterflat images are used to identify and trace all echelle orders for a given night. Once the orders are traced, they are used to extract the spectra of the masterflats. Then, the traces of the orders are used to extract the spectra of the \textit{ThAr} calibration images. Once extracted, an automatized procedure computes the global wavelength solution of the science and calibration fibers of each calibration \textit{ThAr} spectrum. Typically, the number of ThAr emission lines used to compute the wavelength solution is  1\,280 and the root mean square of the residuals is 100 m/s, which translates in a velocity error of 2.5 m/s. The value of 2.5 m/s is the radial velocity floor limit due to wavelength solution. Then, the reference calibration \textit{ThAr} image for a given night is determined by identifying which one produces the smaller bias between the science and calibration fibers in the computation of the instrumental drift which is typically 100-200 m/s during the night, mainly correlating with the atmospheric pressure. A detailed analysis is provided in \citep{fuentes}.

After that, the pipeline uses the optimal extraction algorithm of \cite{2014ascl.soft06002M} to extract the spectra of the science fiber. If the image was obtained with the simultaneous calibration mode, the calibration spectrum is also extracted. The calibration spectrum is used to compute the radial velocity drift with respect to the reference \textit{ThAr} image. The science spectrum is corrected by the blaze function using the extracted \textit{masterflat-sci} spectrum, and then each order is continuum normalized through fitting of a low order polynomial. Finally, the continuum-normalized spectrum is cross-correlated with a binary mask to compute precision radial velocities, bisector span measurements, and full width at half maximum values for the cross-correlation peak. The barycentric radial velocity correction is computed using the routines included into the \texttt{astropy} package \citep{astropy:2013, astropy:2018, astropy:2022}.

\subsection{Data with iodine cell}\label{Sect:I2Cell}

In addition to using a ThAr lamp as a calibration source, PLATOSPec also offers the use of an iodine absorption cell.
There are two options. The first is to use the iodine cell in the classical way and place it in the light path of the observed target (scientific fiber). 
The second option is to observe the iodine cell through the calibration fiber and simultaneously observe the target through the scientific fiber. The second method is similar to the one described in the previous section and is particularly suitable for observing faint targets.

Our iodine cell implementation is designed to prove the new concept and compare the cell usage with the standard ThAr calibration. The average life-time of a ThAr-lamp is between 450 to 1120 hours. Therefore, the ThAr lamp needs to be replaced regularly which requires monitoring of any eventual offsets between old and new lamp. A non-trivial solution of the problem is presented in \citep{2018SPIE10702E..0WQ}. However, our iodine cell calibration modes offer a relatively easy and stable solution with constant calibration lines over years. The limited wavelength range from 520-620 nm needs to be acceptable and the typical iodine cell absorbs about $30\%$ of light. To counter the absorption of light from target, we introduced the simultaneous iodine mode calibration where the cell is lit by the white light source without absorbing the stellar light.

To obtain the RVs from observations taken with the iodine cell, we use the RV pipeline {\tt viper} \citep{viper2021,koehler2025}, which is based on the concept of \citet{butler1996}. It uses a forward model that allows a simultaneous calibration of the wavelength and instrumental profile (IP) using the absorption lines of the iodine cell. Various IP models are available in {\tt viper}, such as a simple Gaussian model or a biGaussian model, in order to take possible asymmetries into account.

{\tt viper} is capable of processing any type of 1D spectra and was developed for spectrographs that use a gas cell as a calibration source. Currently, the observed spectra are pre-reduced with the IRAF package to extract 1D spectra. This pre-reduction consists in the typical steps of basic spectrosccpic data reduction (flat fielding, bias, spectrum extraction). In the future, it is planned to use the {\tt ceres+} pipeline instead.

\begin{figure}
\includegraphics[width=0.495\textwidth]{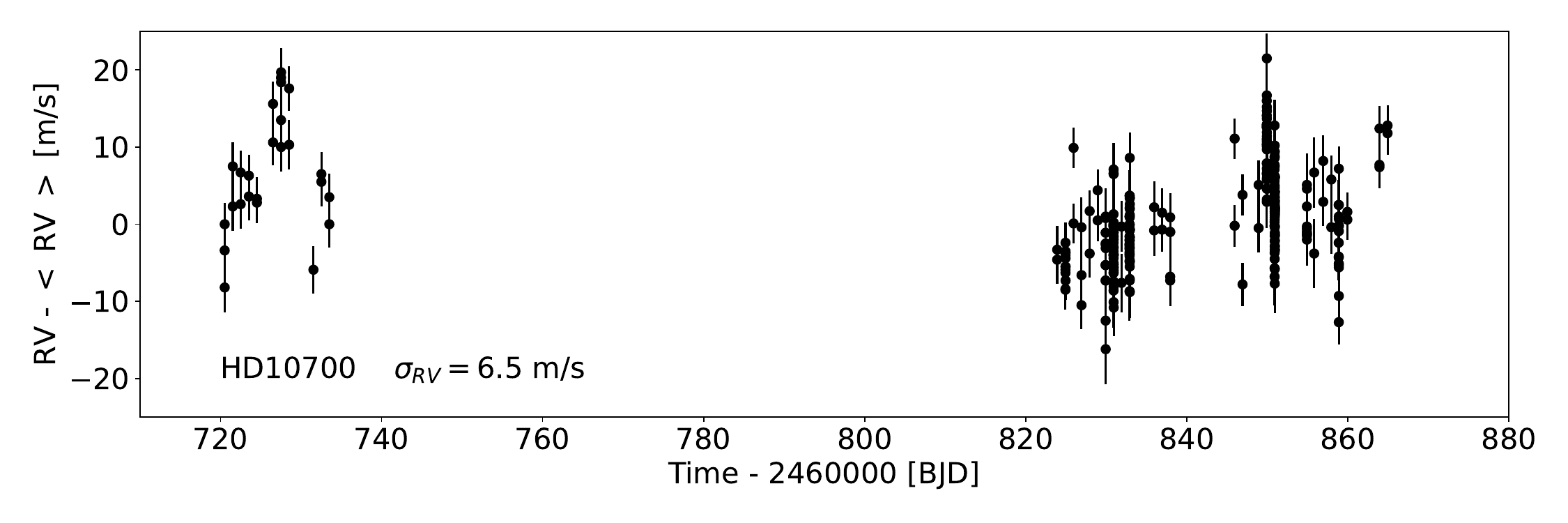}
\includegraphics[width=0.495\textwidth]{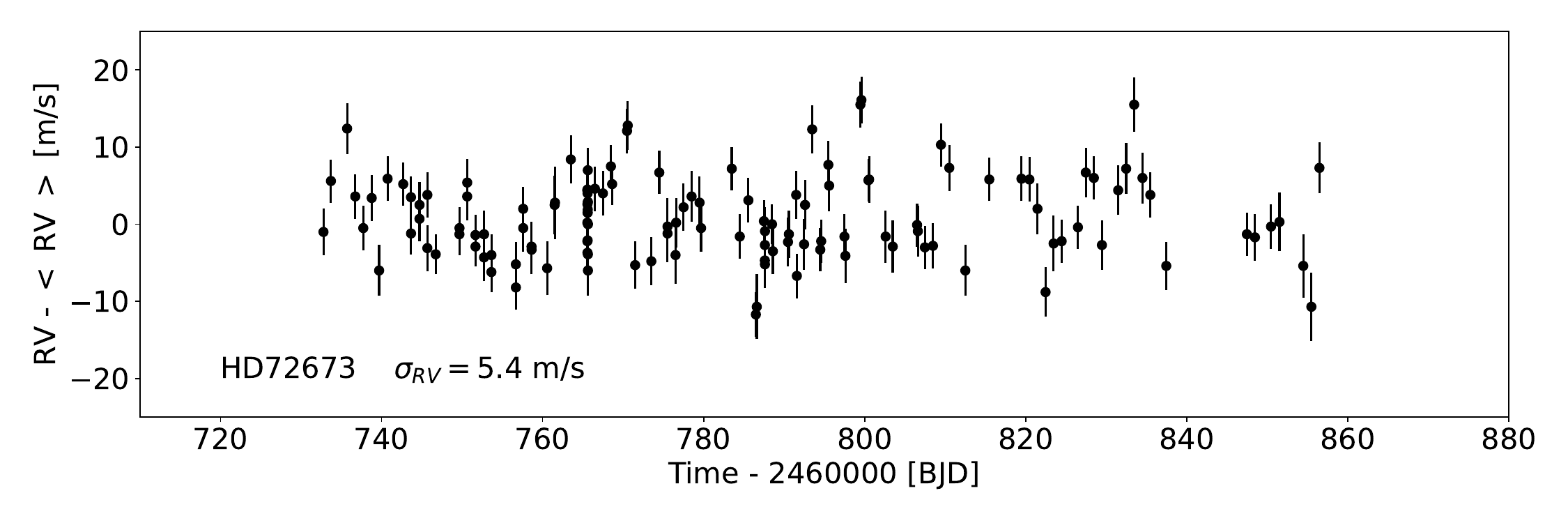}
\caption{Radial-velocity measurements of standard stars during the period of four months of 2025.}
\label{fig:stdrvs}
\end{figure}

\section{The first results from science verification}\label{Sect:FirstResults}

We performed a number of scientific programs within the science verification period since December 2024. Here, we present a few of our very first scientific results, which also demonstrate the capability of the PLATOSpec spectrograph. 


\subsection{Radial Velocity precision with simultaneous ThAr calibration}\label{Subsect:RVprecision}

\begin{figure}
\includegraphics[width=0.495\textwidth]{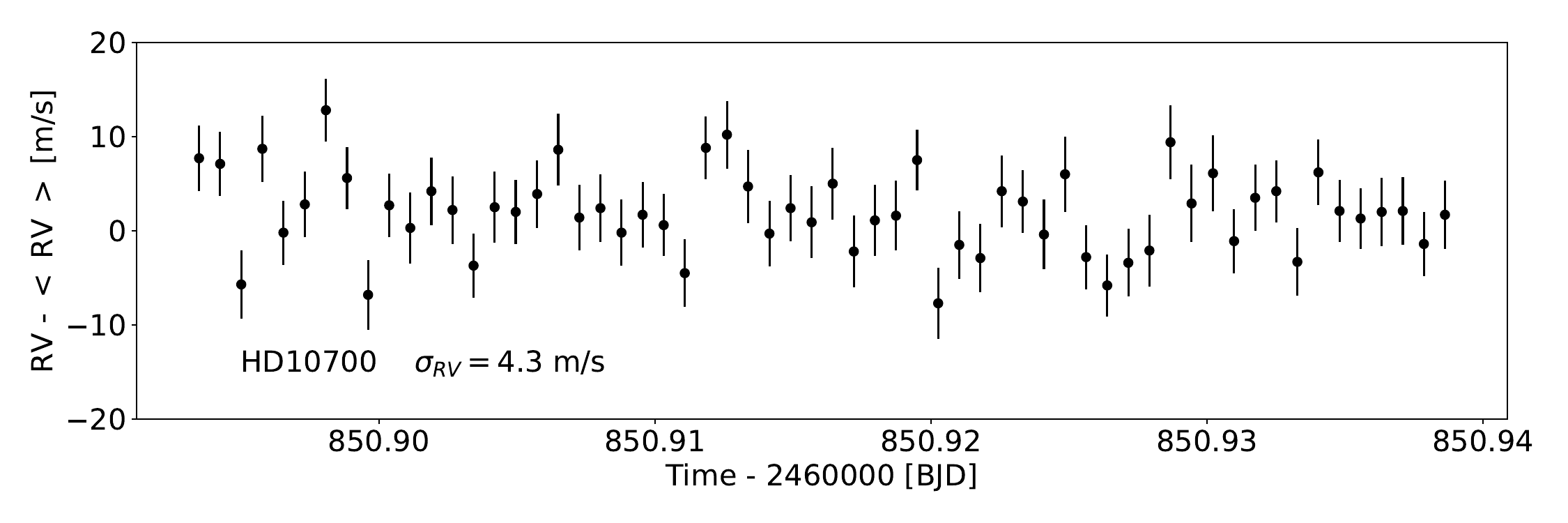}
\caption{Radial-velocity measurements of the standard star HD\,10700 over a period of 1.2 hours.}
\label{fig:stdrvsintra}
\end{figure}

The default calibration mode is the simultaneous ThAr lamp calibration through the ThAr source lamp installed in the calibration unit. A reasonable assessment of the radial velocity (RV) precision and stability of the PLATOSpec instrument can be obtained from observing radial-velocity standard stars using this mode. The measurements of HD\,10700 ($\tau$ Ceti - V$=3.5$\,mag) and HD\,72673 ( V=$6.4$\,mag), spanning from February 2025 until June 2025 are shown in Fig.\,\ref{fig:stdrvs}. The RV stability measured by the standard deviation $\sigma$ is 6.5\,m/s and 5.4\,m/s over a 4 month period for HD\,10700 and HD\,72673, respectively. Furthermore, Fig.\,\ref{fig:stdrvsintra} presents an example of an intra-night variability examined on a sequence of HD\,10700. The scatter over a period of 1.2 hours is 4.3 m/s.

\subsection{Radial Velocity precision with the iodine cell}\label{Subsect:RVsWithI2}

To test the capabilities of PLATOSpec when using the iodine cell, we observed in this mode three bright RV standard stars that show little RV scatter in other surveys. We focus on observations where the iodine cell was placed in the light path of the star. 

\begin{figure}
\includegraphics[width=0.495\textwidth]{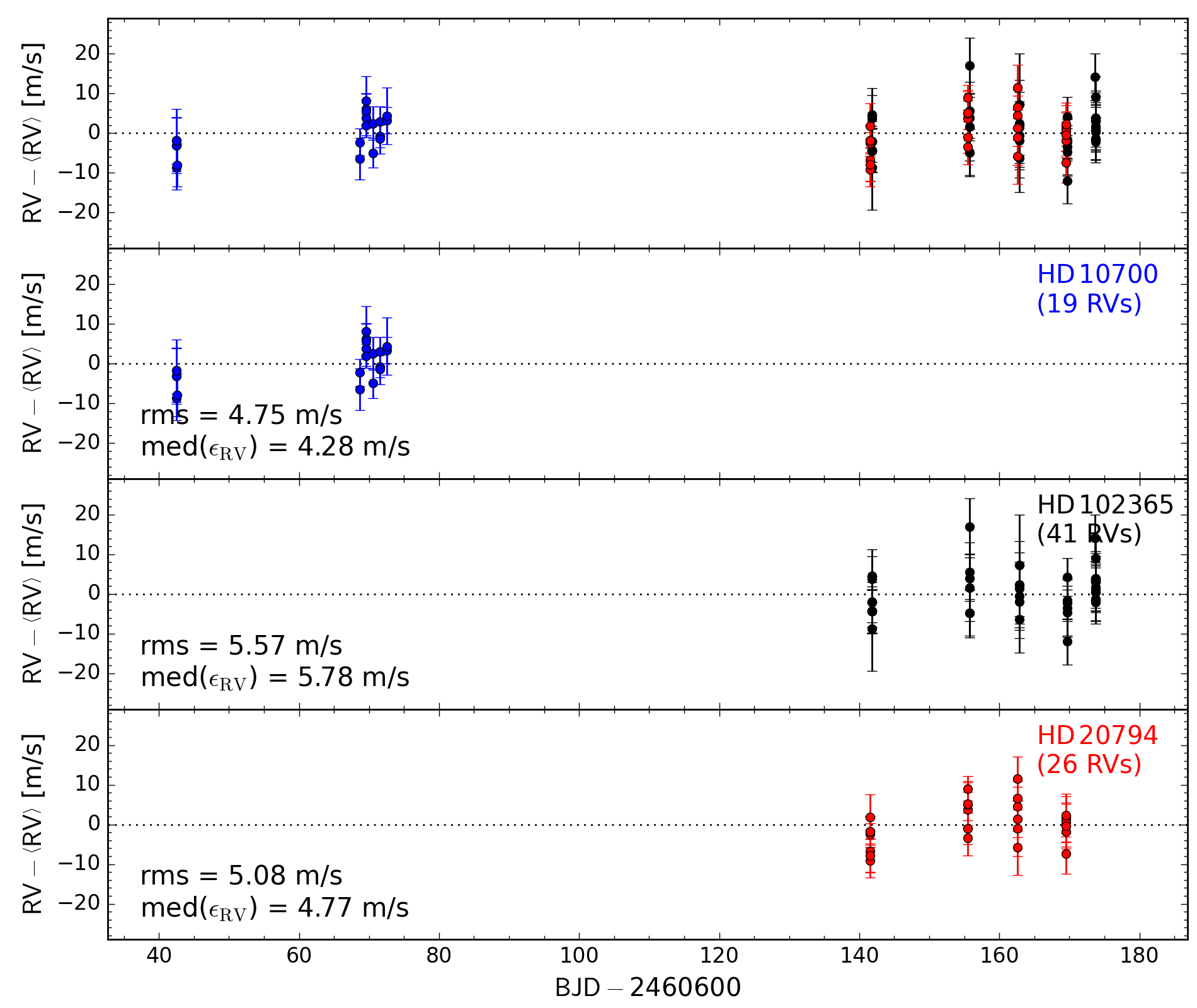}
\caption{RV measurements of the standard star sample with the iodine cell. The med(eRV) corresponds to median error of the radial velocities.}
\label{fig:RV_cell}
\end{figure}

The RVs shown in Fig.~\ref{fig:RV_cell} were obtained with {\tt viper} using a biGaussian IP model, a third degree polynomial for the wavelength solution, and a fifth degree polynomial for the blaze function. Telluric lines were masked to a transmission of 0.98. High-resolution observations from HARPS (for HD\,10700 and HD\,20794 V=$4.3$\,mag) and ESPRESSO (for HD\,102365 V=$4.9$\,mag) were used for the stellar reference templates. The exposure times were set to 120  seconds for HD102365 and $\tau$ Ceti and 180 seconds for HD20794. To account for order-dependent wavelength and IP variations, each echelle order is processed separately. The final RV values are the weighted mean of the RVs from all selected orders \citep[][]{koehler2025}.

For our sample of bright stars, we achieved an RV precision of about 5~m/s over a period of one month. The best results were obtained for the 19 observations of HD\,10700 with an rms of 4.8~m/s. For the 26 RVs of HD\,20794 we calculated an rms of 5.1~m/s and for the 41 RVs of HD\,102365 an rms of 5.6~m/s. The photon noise values for each of the stars are 3.0 m/s for HD102365, 2.4 m/s for HD20794 and 2.0 m/s for HD10700 ($\tau$ Ceti). For all three targets, we found a large inter-night scatter between 4 and 7~m/s. When binning the RVs of one night, the rms between the nights is reduced to 3~m/s.

\begin{figure}
\includegraphics[width=0.495\textwidth]{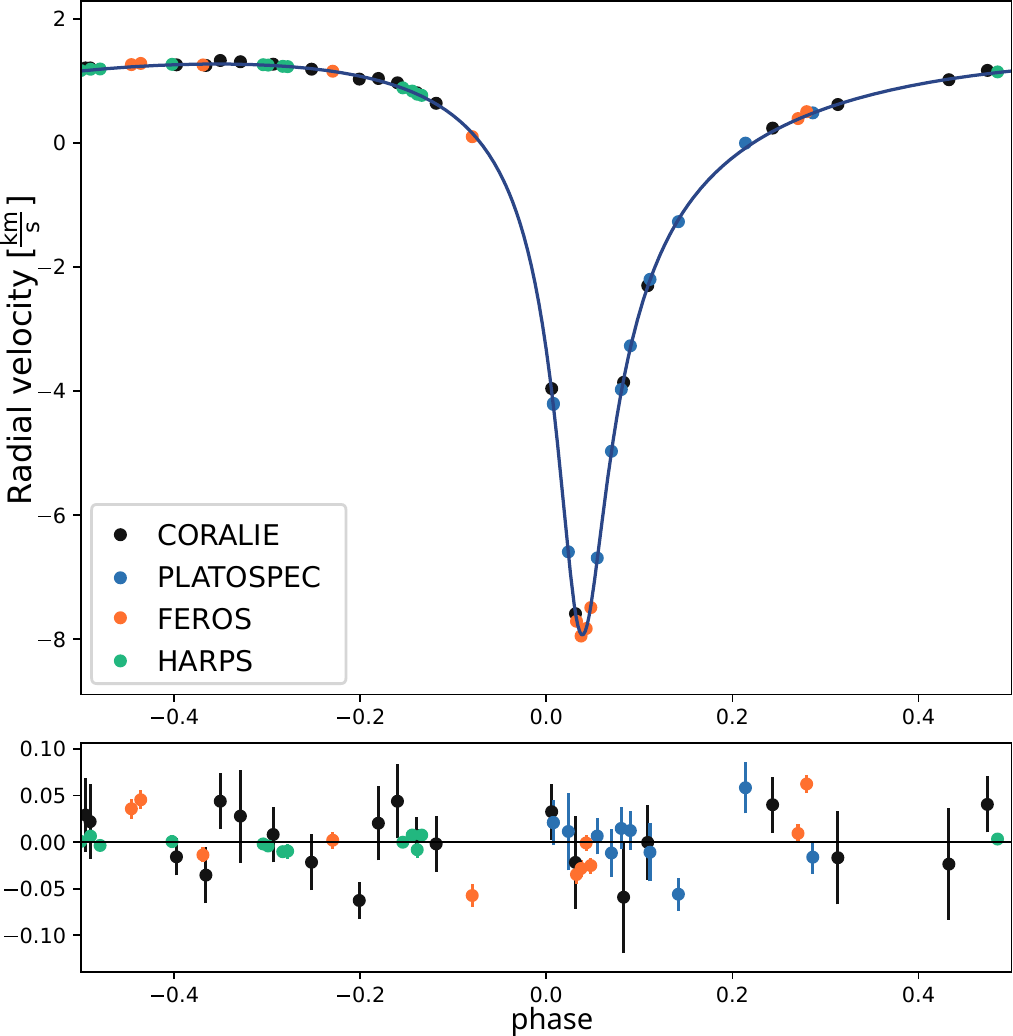}
\caption{Radial velocity curve of TIC\,238060327, combining literature measurements by CORALIE, FEROS, and HARPS, and our new points taken with PLATOSpec. The bottom plot shows the residuals of the best fit model that demonstrate the different precisions and accuracies of the instruments.}
\label{fig:tic}
\end{figure}

\subsection{Eclipsing binaries with low-mass secondary components}\label{Subsect:EBs}
Among the exoplanetary candidates, there are a lot of systems that turn out to be astrophysical false positives. The radius-mass relation of objects in the range between 0.3 -- 100 Jupiter masses ($M_{\rm J}$)  is flat with radii around 1 Jupiter radius \citep{2015ApJ...810L..25H}.  Because the stellar parameters used in the first screening of photometry for potential exoplanetary candidates have rather large errors, it is common for eclipsing binaries with late M-class secondaries or brown dwarfs to be included in the candidate lists. 

Here we present our measurements of TIC\,238060327 \citep[NGTS-EB-7;][]{2025MNRAS.537...35R} demonstrating the ability of PLATOSpec on a rather faint target with $V=12.5$\,mag. In good weather conditions (seeing less than 1 arcsecond, no wind), we were able to achieve a SNR of 26 (at 512 nm) in a 1800\,s exposure, yielding an error on the radial velocity measurements of only 18\,m/s. We note that the host star is of the same spectral type as the cross-correlation mask used and has negligible rotation. We can compare these to the RVs obtained by FEROS (1200s, SNR=60, eRV=9\,m/s) and CORALIE (2100s, SNR=15 at 512 nm, eRV=25\,m/s) used in \citet{2025MNRAS.537...35R}.
Our measurements are listed in Tab.\,\ref{TIC23806032RVs} in the Appendix.

We fit the radial velocities from the four instruments (omitting one outlying HARPS point) using Phoebe~2.4 \citep{2020ApJS..250...34C}. We fixed the ephemeris according to the \citet{2025MNRAS.537...35R} value, which is constrained by three primary eclipses of the system. The resulting fit can be seen in Fig.~\ref{fig:tic}. The data points obtained with PLATOSpec only lead to a marginal improvement of orbital parameters (Table \ref{table:tic}) as these were already well constrained by FEROS (for the semi-amplitude) and HARPS (for  the eccentricity) measurements. 

From the jitter of radial velocities $\sigma_i$ we see that the floor of the stellar contribution is below the HARPS jitter of $\sigma_{HARPS}<4\,\mathrm{m/s}$, CORALIE is limited by the photon noise limit below $\sigma_{CORALIE}<30\,\mathrm{m/s}$.
For FEROS the quoted RV errors are clearly underestimated with jitter of $\sigma_{FEROS}=30\pm14\,\mathrm{m/s}$, given the SNR of the observed spectra and spectral range of the instrument we suppose that this jitter is related to the instrument stability. For PLATOSpec we get $\sigma_{PLATOSpec}<35\,\mathrm{m/s}$ driven mostly by 2 outliers out of the 10 points.
However, the general agreement demonstrates the capabilities of PLATOSpec in the faint target regime.
\begin{table}
\caption{Radial velocity parameters of TIC238060327}        
\label{table:tic}      
\centering     
\begin{tabular}{l | c c}       
Parameter & combined & Rodel 2025 \\ 
\hline
$K$ [km/s] &4.604$\pm$0.010&4.615$\pm$ 0.010\\
$e$ &0.7140$\pm$0.0006&0.7144$\pm$ 0.0009\\
$\omega$ [deg]&170.34$\pm$0.16&170.54 $\pm$ 0.20\\
\hline
$\gamma_{\rm CORALIE}$ [km/s]&78.423$\pm$0.008&78.427 $\pm$ 0.009\\
$\gamma_{\rm FEROS}$ [km/s]&78.410$\pm0.013$&78.418$\pm$ 0.010\\
$\gamma_{\rm HARPS}$ [km/s]&78.317$\pm$0.003&78.314$\pm$ 0.006\\
$\gamma_{\rm PLATOSpec}$ [km/s]&78.430$\pm$0.011&--\\
\hline
$\ln (\sigma_{\rm CORALIE}/\frac{\mathrm{km}}{\mathrm{s}})$ &$-5.1^{+1.3}_{-3.3}$&$-3.99^{+0.4}_{-0.7}$\\
$\ln (\sigma_{\rm FEROS}/\frac{\mathrm{km}}{\mathrm{s}})$ &$-3.3 \pm 0.3$&$-3.5\pm 0.3$\\
$\ln (\sigma_{\rm HARPS}/\frac{\mathrm{km}}{\mathrm{s}})$ &$-7.5^{+1.6}_{-1.8}$&$-4.27\pm 0.3$\\
$\ln (\sigma_{\rm PLATOSpec}/\frac{\mathrm{km}}{\mathrm{s}})$ &$-4.3^{+0.7}_{-3.0}$&--\\
\end{tabular}
\end{table}

\subsection{Mass of exoplanet WASP-79b}\label{Subsect:WASP-79mass}

The exoplanet WASP-79b was discovered by \cite{Smalley2012} using ground-based photometric (WASP and TRAPPIST) and RV measurements (CORALIE). 
This hot Jupiter orbits an F5-type star with $V=10.04$\,mag and a projected rotational velocity of $v \sin i=19\,\mathrm{km/s}$. The orbital period is about 3.66\,days.

We obtained 15 spectra during December 2024 with an exposure time of 1200\,s -- the mean SNR is about 65 (at 512 nm). Using simultaneous ThAr calibration, we were able to measure the RVs with an average uncertainty of 50\,m/s -- a relatively high value due to the rather fast rotation of the star.   Our measurements are listed in Tab.\,\ref{w79} in the Appendix.
We performed an analysis of our RV measurements using the \texttt{juliet} package \citep{juliet}. 
We adopted the published linear ephemeris calculated from transit photometry by \citet{Kokori2023} and further assumed a circular orbit.

\begin{figure}
\includegraphics[width=\columnwidth]{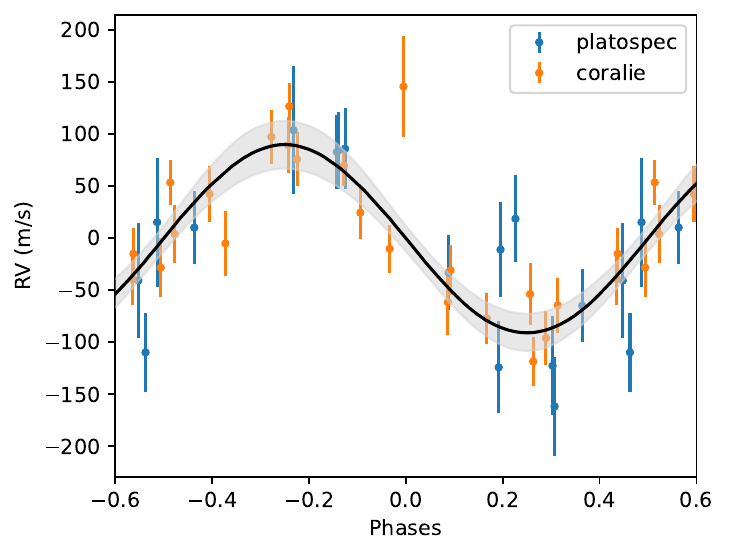}
\caption{Comparison of the PLATOSpec and CORALIE \citep{Smalley2012} RV measurements of the system Wasp-79. The black line shows our model together with the $1\sigma$ interval (grey area).}
\label{fig:wasp79}
\end{figure}

Figure \ref{fig:wasp79} compares our RV measurements with the ones obtained by CORALIE \citep{Smalley2012}. Using only PLATOSpec data, we derived a fit very similar to the one published. The measured RV semi-amplitude is $91.0\pm16.5$\,m/s (compared to a literature value of $88.2\pm7.8$\,m/s). Assuming the same orbital inclination $85.4\pm0.6\degr$ as \citet{Smalley2012}, we find the mass of WASP-79b to be $0.93\pm0.17$\,M$_{\rm J}$. The value is in very good agreement with the mass determined by \citet[$0.90\pm0.09$\,M$_{\rm J}$]{Smalley2012}.

\subsection{Spin-Orbit Alignment of WASP-62b via the Rossiter-McLaughlin Effect}\label{Subsect:RMeffect}

The availability of high-resolution spectrographs has led to an increase of projected spin-orbit angle measurements \citep{bou23,knu24, espi25, zak25a}.
During a planetary transit, the Rossiter-McLaughlin (RM) effect appears as a spectroscopic signature in radial-velocity measurements. This anomaly originates from the selective blocking of light from the rotating stellar surface by the transiting planet \citep{ross24,mcl,alb22}. The analysis of this effect enables the inference of the projected spin-orbit angle ($\lambda$) that measures the alignment between the stellar rotation axis and the planet's orbital axis. This parameter can provide hints on the pathways planets take during their formation and migration. Well-aligned and circular orbits generally point to quiescent migration within the protoplanetary disk, whereas highly misaligned or even retrograde orbits, especially those with significant eccentricity, suggest a history involving dynamical events, such as planet-planet interactions or the Kozai-Lidov mechanism driven by distant companions \citep{fab07}.

To demonstrate the capability of PLATOSpec, we focus here on WASP-62b, a canonical hot Jupiter transiting the F-type star WASP-62 ($V=10.2$\,mag) every 4.4 days \citep{hel12}. Observations of WASP-62 were obtained with PLATOSpec shortly after its installation on the E152 telescope. We observed a single transit of WASP-62\,b on the night of December 1, 2024. The exposure times were between 600-1200\,s and the spectra reached SNRs between 14 and 30 at 512\,nm. This corresponds to radial-velocity uncertainties spanning 34 to 71\,m/s, with a median uncertainty of 44\,m/s. Our measured radial velocities are presented in Tab.\,\ref{w62} in the Appendix.

To derive the projected spin-orbit angle, $\lambda$, of WASP-62b, we employed the approach described in \citet{zak24a}. Our analysis involved fitting the radial-velocity time series with a composite model combining the planet's Keplerian motion with the RM velocity anomaly. This was implemented using the \texttt{ARoMEpy} package \citep{seda23}, which integrates the \texttt{Radvel} module \citep{radvel} for Keplerian modelling and applies the RM formalism from the \texttt{ARoME} code \citep{bou13}.  We set Gaussian priors using  literature values \citep{saha24} and uncertainties derived from transit modelling on the following parameters: the central transit time ($T_C$), the orbital inclination ($i$), and the scaled semi-major axis ($a/R_{\rm{s}}$).  On $\lambda$ and $v\,\mathrm{sin}i_*$, however, we placed uniform uninformative priors.

\begin{figure}
\includegraphics[width=0.495\textwidth]{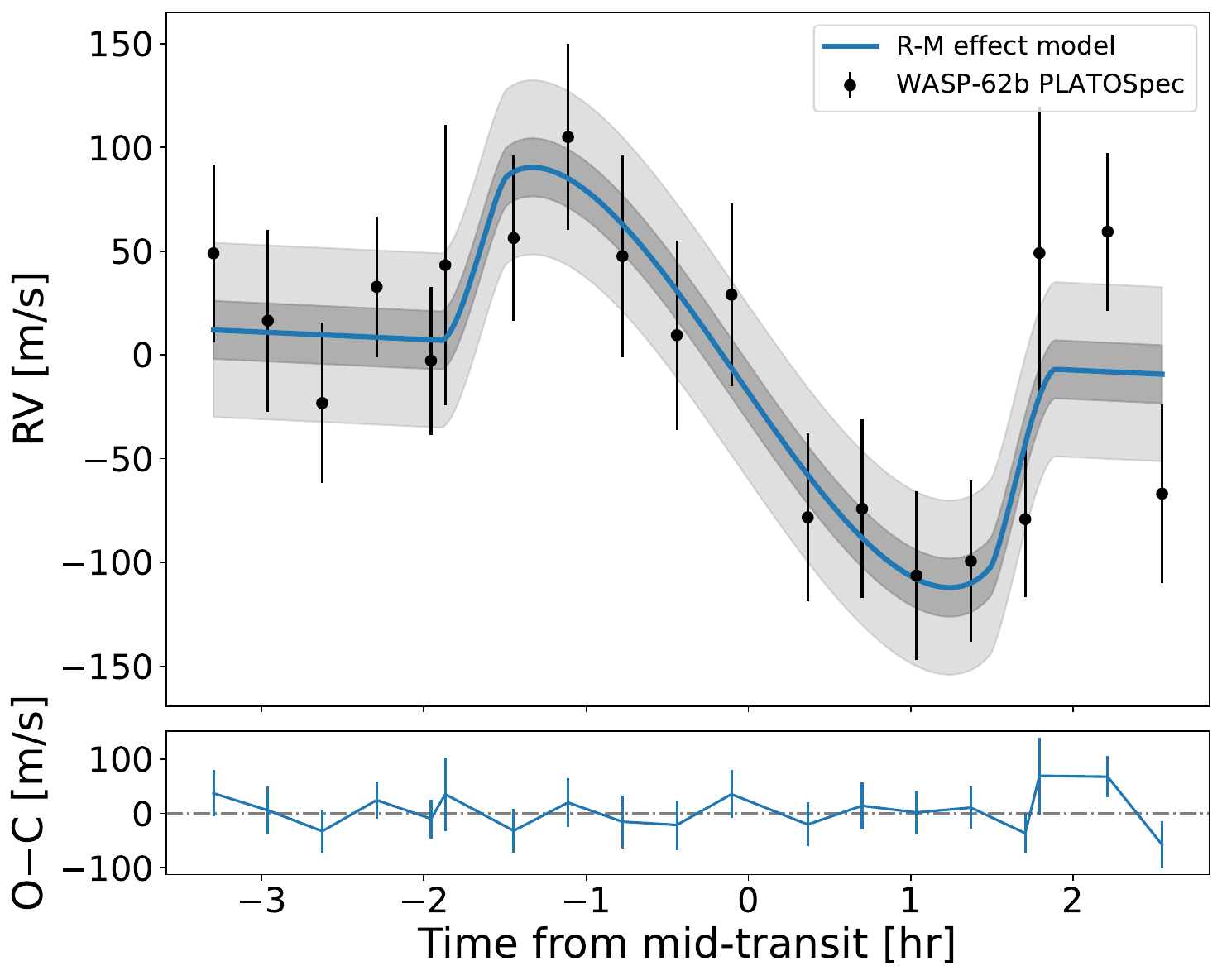}
\caption{The Rossiter-McLaughlin effect of WASP-62b. The observed data points (black) are shown with their error bars. The systemic velocity was removed. The blue line shows the best fitting model to the data, together with 1-$\sigma$ (dark grey) and 3-$\sigma$ (light grey) confidence intervals.}
\label{fig:rme}
\end{figure}

Our MCMC analysis returns a best-fit projected spin-orbit angle of $\lambda =23^{+22}_{-33}$\,deg (Fig. \ref{fig:rme}). This measurement indicates that WASP-62b occupies a prograde orbit relative to the star's equator and is in good agreement with the result $\lambda =18.9^{+11.5}_{-6.6}$\,deg obtained by \citet{br16} using the HARPS instrument mounted on the 3.6-m telescope.

Our measurements of WASP-62\,b add to the evidence demonstrating the utility of instruments like PLATOSpec for RM effect studies. The operational flexibility 
enables the intensive, multi-night observations required to robustly capture the RM signal using a 1.5-m aperture. As ongoing projects like the extended \textit{TESS} survey and upcoming missions such as PLATO continue to identify numerous gas giants transiting bright host stars, instruments with PLATOSpec-like capabilities are well-positioned to perform initial reconnaissance of spin-orbit angles to obtain the initial assessment and identify the most compelling cases (e.g., those potentially misaligned) for dedicated follow-up observations using larger, high-resolution facilities \citep{zak25ps}. Additionally, the spin-orbit angle provides a crucial connection to interpreting the composition of exoplanetary atmospheres \citep{turr21, kirk24}. Upcoming atmospheric missions such as \textit{Ariel} \citep{tin18} will measure the composition of exoplanetary atmospheres for several hundreds of exoplanets, but the spin-orbit angle measurements are currently missing for more than 70\,\% of the potential targets \citep{zak25b}.

\subsection{Line-profile variations in chemically peculiar star $\alpha$\,Cir}\label{Subsect:StelVar}

We made observations of the bright rapidly oscillating Ap star \citep[roAp;][]{Kurtz1982} $\alpha$\,Cir ($V=3.19$\,mag; Simbad spectral type A7VpSrCrEu). This class of stars pulsate in high-overtone low-degree pressure modes with periods on the order of minutes \citep{Kurtz1982,Holdsworth2021,Holdsworth2024}. They show unusual chemical composition due to the radiative lifting of elements with large cross sections, such as those from the iron group and rare-Earth elements \citep{Michaud1970,Preston1974}. Moreover, these rare stars have strong globally-organized magnetic fields that stabilise chemical spots on their surface, leading to rotational modulation \citep[e.g.,][]{Mathys2017,Holdsworth2024} and changes in the strengths and profiles of the spectral lines during a rotation cycle \citep[e.g.,][]{Kochukhov2001,Kochukhov2004}.

\begin{figure}
\includegraphics[width=0.495\textwidth]{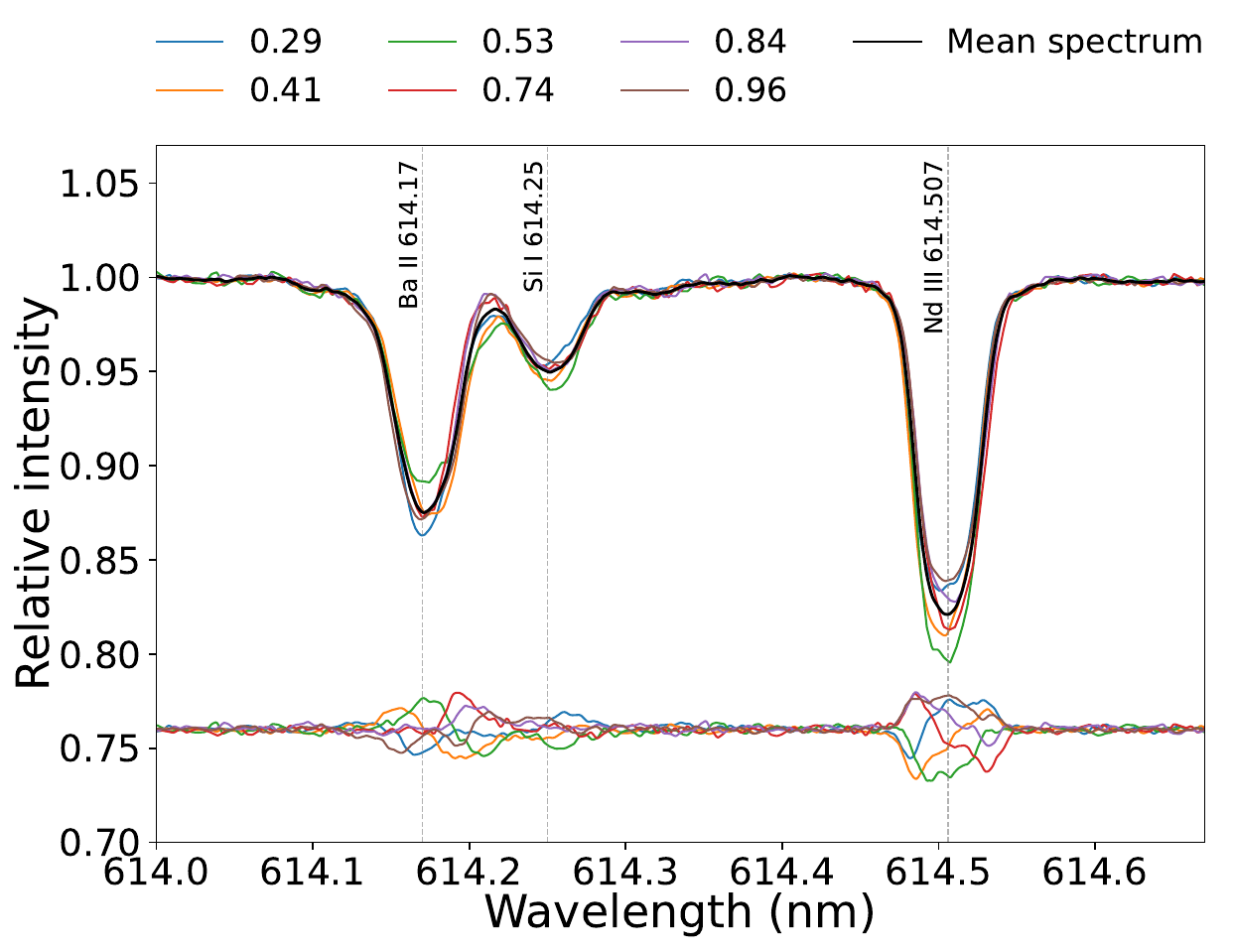}
\caption{Variations of the line profiles of Ba\,\textsc{ii}, Si\,\textsc{i}, and Nd\,\textsc{iii} lines over the rotational cycle. The different colours correspond to different rotation phases (see top legend). The bottom lines show the residual spectra after subtracting the mean spectrum (black line).}
\label{Fig:AlpCirSpec}
\end{figure}

We observed $\alpha$\,Cir at six different rotation phases during eleven nights, each time obtaining series of 13 to 150 spectra (exposure time 30\,seconds, with a median SNR of about 280 over the whole spectrum). We extracted only the spectral order that contains the Ba\,\textsc{ii} 614.17\,nm, Si\,\textsc{i} 614.25\,nm and Nd\,\textsc{iii} 614.51\,nm lines, which show significant variations \citep{Kochukhov2001}. After barycentric RV correction, the spectra were combined to get a mean spectrum for each rotation phase, as well as to get the global mean spectrum from all available spectra. The combination of the spectra ensures that any possible line-profile variations due to pulsations are smeared out. Subsequently, we subtracted the spectra corresponding to the different rotation phases from the mean spectrum. The results are shown in Fig.~\ref{Fig:AlpCirSpec}. As can be seen, there are clear variations in the Ba\,\textsc{ii}, Si\,\textsc{i} and Nd\,\textsc{iii} lines. The Si\,\textsc{i} and Nd\,\textsc{iii} lines change in the same way, while the Ba\,\textsc{ii} line varies differently.

\begin{figure}
\includegraphics[width=0.495\textwidth]{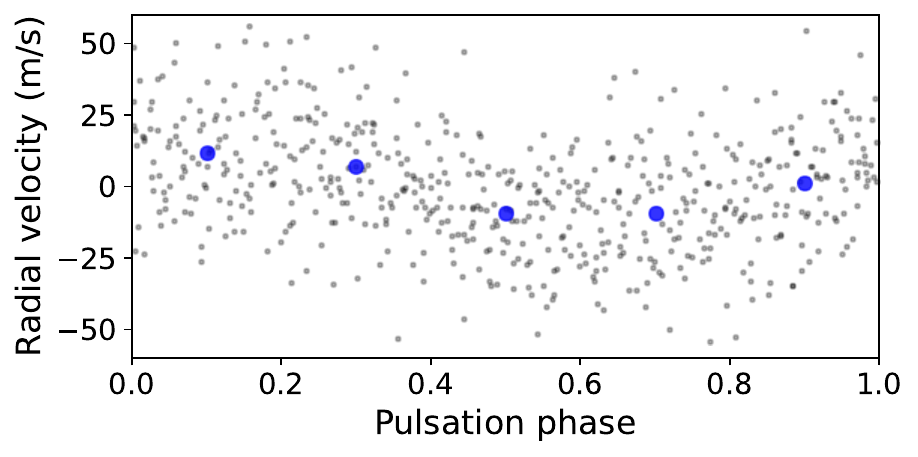}
\caption{Mean radial velocities of $\alpha$\,Cir after removing the rotational variations, showing only the contribution from oscillations. The blue points show the binned data.}
\label{Fig:AlpCirRV}
\end{figure}

On the other hand, we were unable to detect unambiguous line-profile variations of the Ba\,\textsc{ii}, Si\,\textsc{i} and Nd\,\textsc{iii} lines during the pulsation cycle with a frequency of 2442\,$\mu$Hz (210.99\,c/d) \citep{Kurtz1993} that were detected by \cite{Kochukhov2001}. We only detected a barely noticeable excess in the residuals around the Nd\,\textsc{iii} line that is not significant. However, the pulsations are clearly apparent in the mean radial velocities after we subtracted variations with amplitude of $\approx 36$\,m/s, which are connected with the rotational cycle (Fig.~\ref{Fig:AlpCirRV}). The amplitude connected with the mean RV due to the oscillations is about 12.6\,m/s. 

The mean RV error of a single point is about 20\,m/s. The scatter in Fig.~\ref{Fig:AlpCirRV} and the large errors are likely due to the fact that different spectral lines lead to different radial velocities as first established in \citet{1998ApJ...503..848K} and later elaborated in \citet{Baldry1998}, \citet{Balona2003}, \citet{2003MNRAS.345..781M}, and \citet{2008A&A...490.1109M}. A detailed investigation of the pulsations or line profile variations and physical interpretation is beyond the scope of this paper. The goal of this simple exercise with $\alpha$\,Cir was to demonstrate the ability of the PLATOSpec and the \texttt{ceres+} pipeline to produce data appropriate for studying line profile variations and stellar oscillations.

\subsection{UY Pic, a young active star in the southern PLATO field}

UY Pic is an active star in the PLATO field LOPS2. It is a visual binary star consisting of a K0~V primary (HIP~26373; $V$=7.7 mag) and a K6~V secondary (HIP~26369; $V$=9.8 mag), separated by 6.92 arcseconds \citep{2023A&A...674A...1G}. According to Gaia, the distance of the primary is $24.5962\pm0.0083$ pc while that of the secondary is $24.535\pm0.024$ pc \citep{2023A&A...674A...1G}. Both components are very active and the primary is listed in SIMBAD as a RS CVn star. 

\begin{figure}
\includegraphics[width=0.495\textwidth]{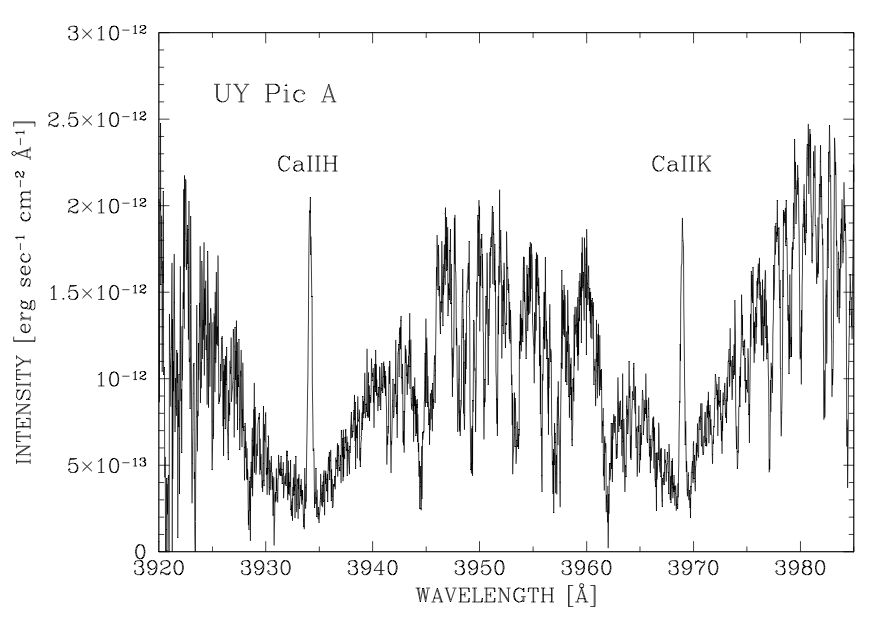}
\caption{PLATOSpec spectrum of part of the flux-calibrated spectrum of UY Pic A showing the Ca~\textsc{ii} H\&K lines.}
\label{Fig:UY}
\end{figure}

We monitored the primary continuously for 5 nights from December 24, 2024 (BJD 2460668.8 to 2460673.6). The spectrum obtained with PLATOSpec around the Ca~\textsc{ii} H\&K region is shown in Fig.~\ref{Fig:UY} and clearly reveals the emission cores in the Ca~\textsc{ii} H \& K lines. We then reobserved the star on January 13, 2025 and January 17, 2025. We find that the RV varied from only 32.63 to 32.80 km/s, which makes it quite unlikely that it is a short-period binary. 
We also observed the secondary on 4 nights between 12 and 16 January 2025 (BJD 2460688, 2460691, 2460689, 2460692). The RV only varied from 31.5 to 33.0~km/s, which also makes it unlikely that it is a short-period binary. 

We analysed the spectra of both stars in detail and measured the equivalent width of the Li~\textsc{i} 6707 line, obtaining $0.2619\pm0.0004$~{\AA} for the primary and an upper limit of $0.021$~{\AA} for the secondary. This is consistent with the results obtained by \cite{2022AJ....164..174H} that identified UY Pic as a member of the AB Doradus moving group with an age of $145^{+50}_{-19}$ Myr \citep{2018ApJ...856...23G}. The large activity  is thus due to their youth, as there no indication that the primary nor the secondary are short-period binary systems.

\section{Conclusions}\label{Sect:Conclusions}

We presented here a new and precise instrument, PLATOSpec, and gave several examples of its potential. PLATOSpec is a high-resolution echelle spectrograph operating in a remote and highly automatized mode. Further improvements towards complete automatization are planned. Furthermore, we also plan to improve the long-term stability and the radial velocity precision. 
Currently, with PLATOSpec 
we achieve a short-term precision of $\approx 4$\,m/s in radial velocities, which degrades slightly to 5-6\,m/s over several months. We also offer simultaneous iodine calibration, improving the precision of radial velocities to $\approx 3$~m/s. 

The PLATO mission will require extensive support for the planetary candidate vetting process, the determination of precise mass and radius of planets, but also for the spectroscopic monitoring of stellar variability. 
PLATOSpec will be a key instrument to provide such ground-based support. The expected SNR values for a sun-like star at 600 nm for a 30 minute exposure and for different apparent brightnesses of $V$=8.5 mag, $V$=11 mag and $V$=13 mag are 82, 25 and 10 with respective radial velocity photon limits of 4 m/s, 13 m/s and 32 m/s (PLATO priority samples, P2, P1, P5). However, PLATOSpec will also follow its own characterisation programs, especially those requiring long time baselines, like the discovery of long-period large planets -- the equivalent of Jupiter in our Solar System.

\section{Data Availability}

The PLATOSpec data are stored in the Ondřejov archive and it is available upon a request. From 2026, the data will be fed into ESO archive where it will be available under the same policy as all other data from ESO instruments. 

\section*{Acknowledgements}
PK, MS and JZ would like to acknowledge the support from the ESA PRODEX under PEA4000127913. We would like to thank Raine Karjalainen for collecting the data in the early phase of the project. We acknowledge travel support from the EXOWORLD project ID 101086149. The research of P.G. was supported by the internal grant No. VVGS-2023-2784 of the P. J. {\v S}af{\'a}rik University in Ko{\v s}ice, funded by the EU NextGenerationEU through the Recovery and Resilience Plan for Slovakia under project No. 09I03-03-V05-00008.
This research has made use of the SIMBAD database, CDS, Strasbourg Astronomical Observatory, France. This work presents results from the European Space Agency (ESA) space mission Gaia. Gaia data are being processed by the Gaia Data Processing and Analysis Consortium (DPAC). Funding for the DPAC is provided by national institutions, in particular the institutions participating in the Gaia MultiLateral Agreement (MLA). 
R.B. acknowledges support from FONDECYT Project 1241963 and from ANID -- Millennium  Science  Initiative -- ICN12\_009.
This research was funded in part by the Austrian Science Fund (FWF) [10.55776/I5711, 10.55776/P37256, 10.55776/PAT4657624].
PLATOSpec was built and is operated by a consortium consisting of the Astronomical Institute ASCR in Ondrejov, Czech Republic (AsU); the Thüringer Landessternwarte (Thuringian State Observatory -- Germany); and the Universidad Catholica in Chile (PUC -- Chile), with some additional minor partners: Masaryk University (Czech Republic), Universidad Adolfo Ibanez (Chile), and the Institute for Plasma Physics of the Czech Academy of Sciences (Czech Republic).  Financing for the modernisation of the 1.52-m telescope was provided by AsU and personal costs were partly financed from grant LTT-20015. Financing for the construction of PLATOSpec was provided by the Free State of Thuringia, represented by the ``Thuringian Ministry for Education, Science and Culture'' as part of the ``Directive for the Promotion of Research''. 
The financial support for the observations is obtained within the framework of the institutional support for the development of the research organization of Masaryk University. The use of the 1.52-m ESO La Silla telescope was made possible through an agreement between ESO and the PLATOSpec consortium.textbf{
We would like to thank to the anonymous referee for a great improvement of the paper.}



\bibliographystyle{mnras}
\bibliography{bibl} 




\begin{appendix}

\section{Effect of atmospheric dispersion on PLATOSpec observations}\label{diffdis}
The refraction of light in Earth's atmosphere causes not only a change in the observed position of stars, but due to the dependence of the index of refraction of air on the wavelength also acts as a dispersion element spreading the starlight in the direction of the parallactic angle. This effect can be examined in the PLATOSpec data sets.

\begin{figure}
\includegraphics[width=0.495\textwidth]{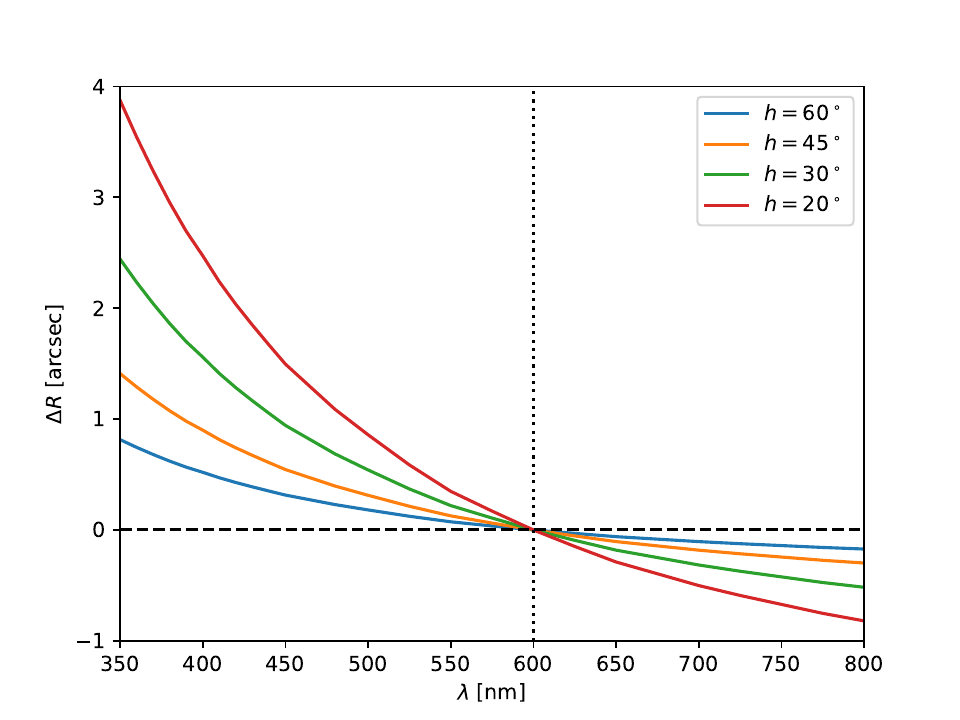}
\caption{
Atmospheric dispersion at the La Silla observatory for different altitudes above horizon of observation $h$ with respect to the reference wavelength of 600 nm where the guiding camera response is the strongest for an early G type star.
}
\label{fig:atmdispt}
\end{figure}

The angle $R$ by which light is refracted can be approximated (when not too close to the horizon) as:
\begin{equation}
    R\approx\left(n_0(\lambda)-1\right)\frac{p}{p_0}\cot h\,,\quad 
\end{equation}
where $n_0(\lambda)$ is the wavelength-dependent index of refraction of air in standard conditions (mainly at atmospheric pressure $p_0$), $p$ is the atmospheric pressure at the observatory\footnote{this factor thus corrects for the altitude of the observatory above sea level} and $h$ is the altitude above horizon of the observed target. For an altitude of $30^\circ$ above the horizon, the difference in $R$ between 400 and 800 nm is about 2\arcsec as can be seen in Fig.~\ref{fig:atmdispt}. The dependence is the steepest in the bluest regions.

\begin{figure}
\includegraphics[width=0.495\textwidth]{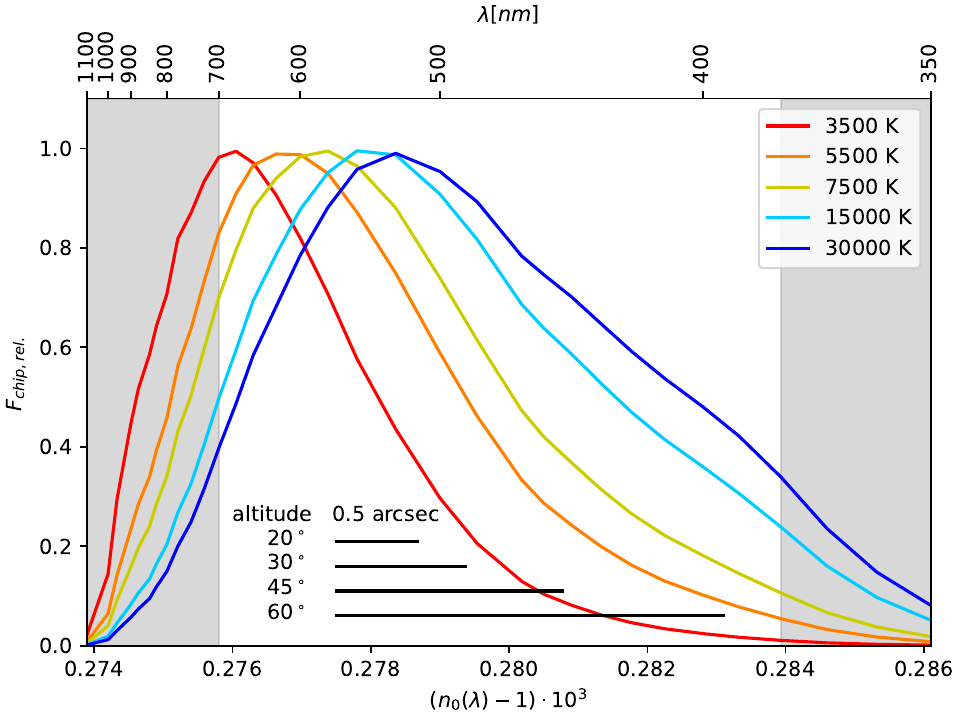}
\caption{Smearing of stars of various effective temperatures on the detector of the guiding camera. The $x$-axis is $n_0(\lambda)-1$, proportional to the spread due to atmospheric refraction, with the corresponding wavelengths noted at the top of the figure. The black lines show the scale of the $x$-axis in arcseconds for different altitudes of observation. These needs to be compared with the usual monochromatic PSFs of the stars (1-1.5 arcsec in good seeing as measured on the guiding images) and the projected fiber entrance pupil with radius of 1 arcsec. The colored lines are normalized blackbody models of different temperature as seen by the guiding camera  (considering the guiding system transmissivity and chip quantum efficiency).  The gray area shows wavelengths outside of the range of PLATOSpec. To estimate the wavelength range injected into the spectrograph for a solar type star observed at altitude of 20 degrees, one takes a window around a maximum of orange curve with a width twice the shortest plotted black line such that the most light falls into it - only wavelengths between 530 nm and 730 nm are injected if we ignore monochromatic PSF, seeing and guiding performance. Even if these contribute as high as 2 arcseconds, we still do not cover the full range of the instrument in the blue.
}
\label{fig:atmdisp}
\end{figure}

Neither the E152 telescope nor PLATOSpec are  equipped with an atmospheric dispersion corrector (ADC) so there is a chromatic loss of light at the fiber entrance. The guiding is performed by Gaussian fit on the observed star-shape on the guiding camera, so the amount by which a specific wavelength is affected depends on the
\begin{itemize}
    \item altitude/airmass of the target during the observation,
    \item spectral energy distribution (SED) of the target,
    \item wavelength in question,
    \item atmospheric seeing, and
    \item stability of the guiding (e.g., due to wind),
\end{itemize}
as we record the convolution of the guiding system sensitivity and the target SED. We note that the effect gets worse with increasing airmass/decreasing altitude as can be seen on Fig.~\ref{fig:atmdisp}. For most of the temperature range of stars observed in the main science program (F,G,K stars, that is, between about 4\,000 to 7\,500~K), the fitted stellar position falls on the uttermost left part of Fig.~\ref{fig:atmdisp}, potentially leaving the blue wavelengths outside of the fiber-entrance, especially in altitudes lower than $45^\circ$ and excellent seeing and cooler stars.

\begin{figure}
\includegraphics[width=0.495\textwidth]{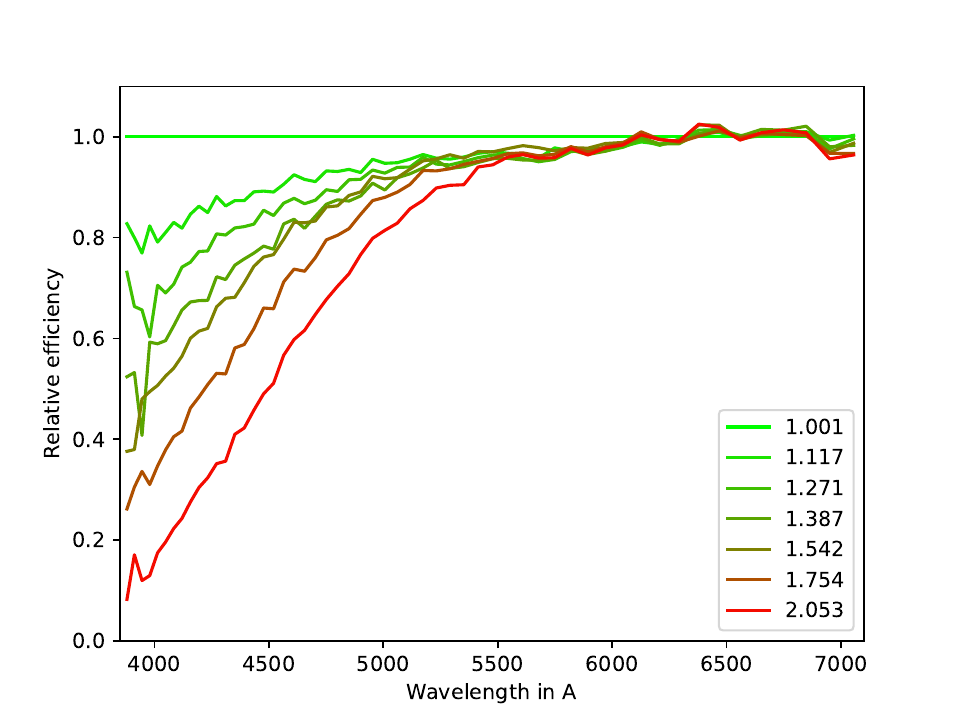}
\caption{Effect of the atmospheric dispersion on observations of a 6\,000\,K star at various airmasses (colored lines).}
\label{fig:ad-obs}
\end{figure}

To demonstrate this effect on real observations, we use observations of TOI-2834 ($T_{\mathrm{eff}}=6000\,\mathrm{K}$, with average seeing FWHM on the guider images of 1.8-2 arcsec during the night, corresponding to average conditions) taken during a series ranging from (almost) zenith down to airmass of 2 (altitude of 30 degrees). Using the per-order median values of extracted fluxes, normalizing them by the zenith observation and scaling by a wavelength independent constant to share a common maximal level, we arrive at Fig.~\ref{fig:ad-obs}. We see that the decrease at the blue end of the spectrograph is noticeable even for relatively small airmass -- at an airmass of 1.5 (altitude of $42^\circ$), we loose 50\% of the light in the bluest part. The flat part of the efficiency relation right of 550 nm is expected from Fig.~\ref{fig:atmdisp} as the effective guiding wavelength for the star is around 600 nm. Given the general decrease of the observed flux in the blue region, this effect contributes only to a small decrease of the SNR of the cross-correlation function and thus only in a slight increase of the radial velocity errors. However, by changing the effective weights of individual lines, the changing altitude can introduce systematic shifts to the measured radial velocities. More impacted are science cases interested in the blue region itself, e.g., the derivation of the activity using the H\&K calcium lines, which should be scheduled as close to the culmination as possible, and counter-intuitively, in worse than average seeing.

\section{Measured radial velocities}\label{measured_RVs}
Below we present the radial velocities measured with PLATOSpec using the simultaneous ThAr calibration mode and the {\tt ceres+} data reduction pipeline, as derived by cross-correlation with a G2V binary mask. These data are presented as some exemplary science cases in Sec.~\ref{Subsect:EBs} to \ref{Subsect:RMeffect}.

\begin{table}
	\centering
	\caption{PLATOSpec radial velocities of the TIC 23806032 system}
	\label{TIC23806032RVs}
	\begin{tabular}{lccc} 
		\hline
		Julian Date (BJD) & RV (km/s) & RV err (km/s) & SNR  \\
		\hline
2460647.7181780	 &74.3010	 &0.0243  &20       \\
2460650.8301193	 &71.9071	 &0.0417  &13       \\
2460659.7748403	 &73.5294	 &0.0253  &18       \\
2460656.8296594	 &71.8092	 &0.0196  &24       \\
2460661.8239382	 &74.5249	 &0.0226  &20       \\
2460663.7089809	 &75.2292	 &0.0208  &21       \\
2460667.7882969	 &76.3011	 &0.0312  &15       \\
2460673.6901053	 &77.2312	 &0.0178  &26       \\
2460687.6183288	 &78.4977	 &0.0275  &16       \\
2460701.6025821	 &78.9854	 &0.0177  &26       \\
        \hline
	\end{tabular}
\end{table}

\begin{table}
	\centering
	\caption{PLATOSpec radial velocities of the WASP-79 system}
	\label{w79}
	\begin{tabular}{lccc} 
		\hline
		Julian Date (BJD) & RV (km/s) & RV err (km/s) & SNR  \\
		\hline
        2460647.65268869  &  4.7781 & 0.0441 &  65 \\
2460647.66663522  &  4.8913 & 0.0456 &  63 \\
2460647.77908704  &  4.9209 & 0.0418  & 71 \\
2460648.59077462  &  4.8616 & 0.0553 &  52 \\
2460648.73236938  &  4.9176 & 0.0619 &  45 \\
2460649.76437611  &  5.0060 & 0.0616 &  45 \\
2460651.72234678  &  4.7796 & 0.0475 &  61 \\
2460651.73748535  &  4.7405 & 0.0476 &  60 \\
2460653.75394601  &  4.9850 & 0.0357 &  85 \\
2460653.76790402  &  4.9868 & 0.0370 &  81 \\
2460653.81600473  &  4.9882 & 0.0390 &  76 \\
2460659.63221446  &  4.7925 & 0.0378 &  80 \\
2460663.66288940  &  4.9125 & 0.0355 &  86 \\
2460665.58602517  &  4.8693 & 0.0365 &  83 \\
2460666.59545177  &  4.8374 & 0.0354 &  85 \\
		\hline
	\end{tabular}
\end{table}

\begin{table}
	\centering
	\caption{PLATOSpec radial velocities of the WASP-62 system}
	\label{w62}
	\begin{tabular}{lccc} 
		\hline
		Julian Date (BJD) & RV (km/s) & RV err (km/s) & SNR  \\
		\hline
        2460646.62621710 & 14.9810 & 0.0428 & 24 \\
2460646.64016385 & 14.9485 & 0.0439 & 24 \\
2460646.65412218 & 14.9088 & 0.0387 & 27 \\
2460646.66808051 & 14.9648 & 0.0339 & 30 \\
2460646.68203884 & 14.9292 & 0.0357 & 29 \\
2460646.68569624 & 14.9753 & 0.0675 & 15 \\
2460646.70324253 & 14.9883 & 0.0400 & 26 \\
2460646.71720086 & 15.0370 & 0.0449 & 23 \\
2460646.73115919 & 14.9796 & 0.0486 & 21 \\
2460646.74511751 & 14.9415 & 0.0457 & 22 \\
2460646.75914528 & 14.9610 & 0.0441 & 23 \\
2460646.77870545 & 14.8537 & 0.0403 & 25 \\
2460646.79266377 & 14.8578 & 0.0430 & 23 \\
2460646.80661052 & 14.8256 & 0.0408 & 25 \\
2460646.82056884 & 14.8326 & 0.0389 & 26 \\
2460646.83452716 & 14.8528 & 0.0374 & 27 \\
2460646.83817299 & 14.9811 & 0.0707 & 14 \\
2460646.85567298 & 14.9914 & 0.0381 & 27 \\
2460646.86963129 & 14.8651 & 0.0430 & 24 \\
		\hline
	\end{tabular}
\end{table}

\end{appendix}
\end{document}